# Full control of Co valence in isopolar LaCoO$_3$ / LaTiO$_3$ perovskite heterostructures via interfacial engineering

Georgios Araizi-Kanoutas[1*a], Jaap Geessinck[2*], Nicolas Gauquelin[3], Steef Smit[1], Xanthe Verbeek[1], Shrawan K. Mishra[4], Peter Bencok[5], Christoph Schlueter[6], Tien-Lin Lee[5], Dileep Krishnan[3], Jo Verbeeck[3], Guus Rijnders[2], Gertjan Koster[2] and Mark S. Golden[1b]

[1]Van der Waals-Zeeman Institute for Experimental Physics, Institute of Physics, University of Amsterdam, Science Park 904, 1098 XH Amsterdam, The Netherlands

[2]MESA+ Institute for Nanotechnology, University of Twente, Faculty of Science and Technology, P.O. Box 217, 7500 AE Enschede, The Netherlands

[3]Electron Microscopy for Materials Science, University of Antwerp, Campus Groenenborger Groenenborgerlaan 171, 2020 Antwerpen, Belgium

[4]School of Materials Science & Technology, Indian Institute of Technology (BHU), Varanasi-221 005, India

[5]Diamond Light Source Ltd, Diamond House, Harwell Science & Innovation Campus, Didcot, OX11 0DE, United Kingdom

[6]PETRA III, DESY Photon Science, Notkestr. 85, 22607 HAMBURG, Germany

*These two authors contributed equally to this work.
[a] G.AraiziKanoutas@uva.nl   [b] M.S.Golden@uva.nl

**PhySH:** Complex oxides; transition metal oxides; X-ray absorption; X-ray magnetic circular dichroism; thin films; scanning transmission electron microscopy.

## Abstract

We report charge-transfer up to a single electron per interfacial unit cell across non-polar heterointerfaces from the Mott insulator LaTiO$_3$ to the charge transfer insulator LaCoO$_3$. In high-quality bi- and tri-layer systems grown using pulsed laser deposition, soft X-ray absorption, dichroism and STEM-EELS are used to probe the cobalt 3d-electron count and provide an element-specific investigation of the magnetic properties. The experiments prove a deterministically-tunable charge transfer process acting in the LaCoO$_3$ within three unit cells of the heterointerface, able to generate full conversion to 3d$^7$ divalent Co, which displays a paramagnetic ground state. The number of LaTiO$_3$|LaCoO$_3$ interfaces, the thickness of an additional 'break' layer between the LaTiO$_3$ and LaCoO$_3$, and the LaCoO$_3$ film thickness itself in tri-layers provide a trio of sensitive control knobs for the charge transfer process, illustrating the efficacy of O2p-band alignment as a guiding principle for property design in complex oxide heterointerfaces.

## Introduction

Complex oxides of the transition metals are of great importance and interest both from a fundamental science as well as a technological point of view. Technologically, LiCoO$_2$ has underpinned the development of the now eponymous Li ion battery [1], ferrites are indispensable in transformer cores and inductors and oxide piezoelectric materials, such as PZT and their Pb-free analogues, are enablers of ultrasound imaging [2,3]. With regards to fundamental science, complex oxides show an interplay between strong electron correlations, band behaviour, as well as rich repertoire of ordering phenomena in the spin and orbital sectors making them an enduring focus of theoretical and experimental investigation [4,5].

The maturity of epitaxy-based thin film growth techniques provides opportunities to improve experimental control over the properties of the system, leading to novel, emergent interfacial properties such as conductivity [6], magnetism [7] and superconductivity [8,9]. A central concept in the field is the role played by nature's response to an incipient polar catastrophe at interfaces, which can be seen as an ultimate driver of interfacial charge transfer [10], providing an elegant and powerful new mechanistic paradigm for doping at a distance in an interfacial system. In practise, the response of real materials to the presence of a potentially polar interface can also be rooted in the relatively facile creation of oxygen vacancies in perovskite transition metal oxides [11], meaning in some systems the in-built potential of a polar overlayer can be observed [12], and in other cases, such as the $LaAlO_3/SrTiO_3$ system it is not [13]. In addition, strain and the $GdFeO_3$-(or octahedra tilting) distortion of the cubic $ABO_3$ perovskite can also be either transmitted or blocked between a bulk substrate and an overlayer [14]. All of these, and other properties constitute a number of tools that can be used to design novel functionalities in complex oxides [15]. As the dominant, silicon-based electronic materials universe makes clear, the interface can indeed be the device [16], fostering added interest in (ultra)thin oxide films and their interfacial properties.

One method to tune and control d-state occupation is via charge transfer, and recently, a broadly applicable principle was introduced enabling the design of oxide heterointerfaces in which charge transfer is predicted [17]. The idea here is that both octahedral backbone and the A-site sublattice in an $ABO_3$ perovskite can be considered continuous across an $ABO_3$-$AB'O_3$ heterointerface. As a consequence, the O2p-related bands of the two materials should align in energy. Depending on the relative energy alignment and separation of the metal (B or B') 3d and O2p states in each compound, this can lead to charge transfer becoming favourable in the heterointerface, also in the case of *isopolar* heterointerfaces. The observation [18] of charge transfer from Ti to Fe at isopolar interfaces between $LaTiO_3$ (LTO) (ground state $3d^1$) and $LaFeO_3$ (ground state $3d^5$) was an important inspiration for the development of the O2p-band alignment picture [17].

In this paper, we report the successful interfacial transfer of a full electron per (lateral) unit cell from LTO to $LaCoO_3$ (LCO), transforming the trivalent $3d^6$ cobaltate into a $3d^7$ divalent state, displaying significant paramagnetic polarisation in external magnetic fields. This charge transfer to a $3d^7$ LCO configuration without any chemical doping or change in the structure is operative at the nanoscale, being concentrated within three unit cells of the LCO/LTO interface, thus allowing tuning of the average Co valence by choice of the LCO thickness and number of interfaces. This achievement is all the more remarkable since the very electronic driving force for charge transfer could easily result in unwanted chemical effects and uncontrolled defect formation. As we will argue below, for the thin layers we are studying, we have ample of experimental evidence that the electronic charge transfer is indeed the leading effect, meaning a transformation from tri- to divalent Co is achieved without changing the structure.

These experimental findings support the design guidelines involving O2p-band alignment for the generation of interfaces displaying nanoscale charge transfer [17]. Nanoscale, controlled electron transfer processes in oxides – in particular in cobaltates - could also be very interesting in the context of sustainable energy technologies involving the oxygen evolution reaction, in which activity has been linked to an $e_g$ electron occupancy of unity [19].

Sample design and fabrication

LCO can be considered the parent compound of many interesting complex cobalt oxides. In its ground state, bulk LCO is a $d^6$ charge transfer insulator with a nominally low-spin, non-magnetic configuration ($t_{2g}^6 e_g^0$). Early temperature dependent studies [20] suggested a gradual transition from low spin (LS) to intermediate spin (IS) states, but more recent studies were clear that the LS ground state co-exists with a triply-degenerate HS excited state, to yield an inhomogeneous mixed-spin state [21]. In the experiments reported here, ultrathin films of LTO and LCO were grown using Pulsed Laser Deposition (PLD) on conducting 0.5 weight percent Nb-doped, (100)-oriented $SrTiO_3$ (STO) substrates. The substrates were ultrasonically cleaned in acetone and subsequently ethanol, followed by a chemical etching procedure and finally annealing to achieve a well-defined, single $TiO_2$-terminated surface [22]. Intensity variations in Reflection High Energy Electron Diffraction (RHEED) were used to monitor the growth and assure unit-cell level control over the film thickness. Subsequently, a 30 unit cell (uc) thick layer of $LaAlO_3$ (LAO) was grown. This served to (a) enable an accurate calibration of the PLD set-up as LAO grows very well in a layer-by-layer manner; (b) prevent formation of a polar-interface-driven charge transfer involving the LCO or LTO (both of which are polar) by separating them far from the non-polar STO, (c) mask the Ti states of the STO substrate in the soft X-ray absorption experiments and (d) block transport of oxygen from the STO substrate into the LTO film [23]. After growth of the LTO, LCO or combinations thereof, a 5 uc $LaNiO_3$ (LNO) layer was added as a cap, so as to enable ambient transfer of the samples to the synchrotron and reduce the chance of charging effects.

The growth of LCO and LTO film combinations poses a dilemma: in order to grow LTO, a low oxygen background pressure (typically well below $10^{-4}$ mbar) is desired, in order to avoid formation of unwanted phases like $La_2Ti_2O_7$ [24] and over-oxidation of the LTO, which results in tetravalent Ti with $3d^0$ electronic configuration [23]. On the other hand, good LCO growth prefers a higher oxygen background pressure (typically ~0.1 mbar), so as to avoid oxygen vacancies [25]. Thus, while interfacing perfect LTO and LCO in the computer [17] is relatively straightforward, in the laboratory, true high-pressure LCO growth would aggressively over-oxidize the underlying LTO, resulting in an uncontrolled LTO quality, and low-pressure growth aimed at stoichiometric LTO will not allow the growth of stoichiometric LCO with trivalent Co. Consequently, a third way was chosen here: namely, to grow both materials at an intermediate pressure of $2*10^{-3}$ mbar. Several strategies were combined to keep the impact on the LTO as minimal as possible, for example by keeping the LTO thickness below 5 uc and utilising the substrate-induced strain to help stabilise the 113-phase of LTO [24]. We note that all reference samples without LTO were still grown at this intermediate pressure, in order to avoid sample variations due to differences in growth conditions. Other PLD-parameters were optimized to obtain a flat and smooth surface after deposition and to form sharp interfaces.

An overview of all the growth parameters can be found in Table I. Below we will experimentally justify the appropriateness of the chosen synthesis conditions.

Table I. PLD growth parameters for the LAO, LTO, LCO and LNO layers. P[$O_2$] denotes the oxygen background pressure during growth.

| Material | Fluence (J/cm$^2$) | Substrate temperature (°C) | Laser rep. rate (Hz) | P[$O_2$] (mbar) | Laser spot size (mm$^2$) |
|---|---|---|---|---|---|
| LaAlO$_3$ | 1.3 | 750 | 1 | 2x10$^{-3}$ | 2.3 |
| LaTiO$_3$ | 1.9 | 750 | 1 | 2x10$^{-3}$ | 2.3 |
| LaCoO$_3$ | 1.9 | 850 | 2 | 2x10$^{-3}$ | 2.3 |
| LaNiO$_3$ | 1.9 | 750 | 1 | 2x10$^{-3}$ | 2.3 |

A wide range of samples were grown, designed to test various aspects of the expected physical behaviour. As mentioned above, all possess a 30 uc LAO buffer layer and a 5 uc LNO capping layer (unless otherwise stated) and can be divided into four differing sample types as depicted in Fig. 1.

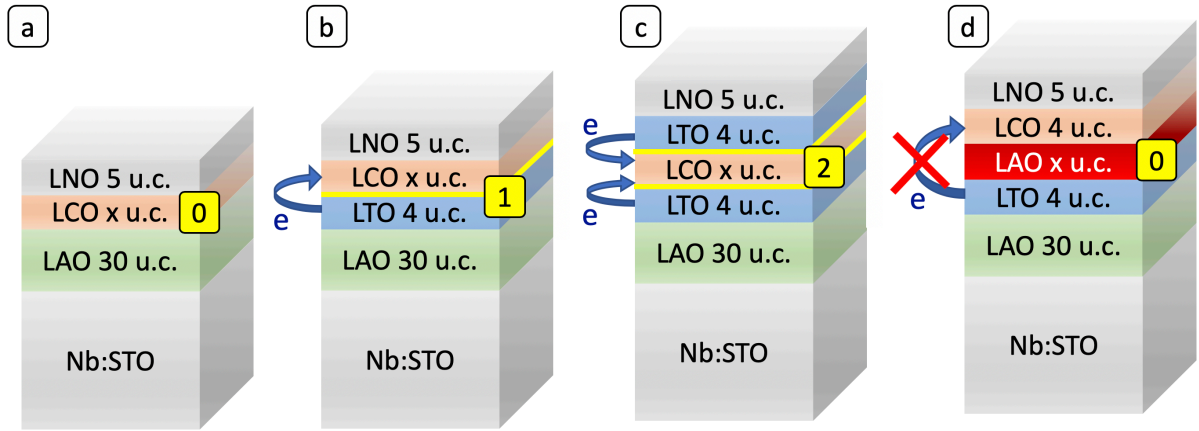

Figure 1: Family of samples according to the number of 'active' LaTiO$_3$|LaCoO$_3$ interfaces [IFs] (yellow highlight): (a) no LTO| LCO IF [0*IF] (b) [1*IF] LTO|LCO (c) [2*IF] LTO|LCO (d) no direct LTO|LCO IF [0*IF] due to LAO 'break' layer. All samples were grown on Nb:STO substrates with a LAO buffer layer of 30 uc and a 5 uc LNO cap.

The four sample types are:
(a) LCO - containing no interface [0*IF]
(b) LTO | LCO - containing a single heterointerface [1*IF]
(c) LTO | LCO | LTO - containing a double heterointerface [2*IF]
(d) LTO | LAO | LCO – a single interface system including a 'breaker' layer of LAO.
To exclude strain-related effects dominating the physics observed a [2*IF] sample was generated on a bulk LAO substrate.
Since the bulk (pseudocubic) lattice constant of LCO is 3.78 Å at 4K and 3.84 at 1248K [26] the LCO would have a tensile strain of about 3% when fully strained on the STO substrates. For the samples grown on bulk (100) LAO substrates, LCO would have a compressive strain of about 1%.

### Soft X-ray absorption, x-ray magnetic circular dichroism and hard X-ray photoemission

X-ray Absorption Spectroscopy (XAS) was carried out at the Co-L$_{2,3}$ (and Ti-L$_{2,3}$) edges using soft X-rays from the i10 beamline at Diamond Light Source, Didcot, in the BLADE end-station. Both Total Electron Yield (TEY) and Fluorescence (FY) detection modes were employed simultaneously. TEY probing depths are ~10nm at the Co-L$_3$ edge [27], and the effective FY probing depth is comparable to the X-ray penetration depth, although FY data do present some complications for the use of XMCD sum rules [28]. Consequently, the XMCD data analysis presented here concerns the TEY data, with the FY data used as a double-check for thicker films. The experimental station combines a cryostat operated at a lowest temperature of 10K with a superconducting magnet applying fields between $-14 \leq H \leq +14$ Tesla. The magnetic field is oriented along the incoming X-ray beam-path, facilitating element specific magnetometry based on x-ray magnetic circular dichroism (XMCD) experiments. The base pressure of the measurement chamber is in the $10^{-10}$ mbar range.

The $L_{2,3}$-edge XAS spectrum of a transition metal compound involves electronic transitions from the $2p_{1/2}$ and $2p_{3/2}$ core levels to unoccupied 3d states, and provides an ultra-sensitive fingerprint of the d-electron count, or valence of the system under investigation [29].

For ensuring maximally accurate XAS data, experiments were conducted at 10K by recording repeated blocks of spectra with alternating X-ray polarisation (e.g. σ+ σ- σ+ followed by σ- σ+ σ-). For the valence fingerprinting, the average of the spectra for the two circular polarizations is taken. For the majority of the data reported here, the X-rays were incident at a grazing angle of 20 degrees with respect to the surface of the film. Control experiments were carried out at higher temperatures, larger incidence angles and investigating different locations on the 5x5 mm$^2$ films. In all cases, the measured XAS spectra were normalized to the edge-jump, accounting for the number of holes in the 3d-shell, while a linear background that was determined well before the pre-edge region was subtracted.

Hard X-ray photoemission measurements were conducted at the I09 beamline at DIAMOND Light Source using a photon energy of 2.2 keV. The HAXPES spectra were recorded using an EW4000 photoelectron analyzer (VG Scienta), equipped with a wide-angle acceptance lens, with the X-rays incident at a grazing angle of 55°, so as to enable depth profiling analysis of the photoelectrons where appropriate. Soft X-ray absorption spectra were also recorded at the I09 beamline at the Co $L_{2,3}$ edges from the same spot (30 μm × 50 μm) as was measured using HAXPES, so as to connect to the XAS data recorded at the I10 beamline. A portable UHV 'suitcase' chamber was used to transfer some of the samples from the PLD system in Twente to DIAMOND in a pressure in the $10^{-10}$ mbar range, to test for oxidation during transport.

The in-situ XPS data shown in the supplementary material (Fig. S4) were recorded in the Twente multi-chamber PLD system using monochromatized Al:Kα radiation and an Omicron 7-channeltron electron energy analyzer.

Scanning Transmission Electron Microscopy

Scanning Transmission Electron Microscopy (STEM) using High Angle Annular Dark Field (HAADF) imaging was performed using a FEI Titan 80-300 microscope operated at 120 kV. The samples were prepared in a vacuum transfer box and studied while held in a Gatan vacuum transfer sample holder to avoid any influence of air on the film [30-32]. Electron Energy Loss Spectroscopy (STEM-EELS) measurements were performed using a monochromatic beam with a 120meV energy resolution. The Ti L, Co L edge, O-K and La-$M_5$ edges were acquired simultaneously (the La being used for energy calibration). The acquisition parameters were 0.25s/pixel, 0.4Å/pixel and 0.05eV/pixel in the dual EELS mode. Collection angles for HAADF imaging and EELS were 70-160 mrad and 47 mrad, respectively.

Results and discussion – XAS data

As the right-hand panel (b) of Fig. 2 shows, Co-$L_{2,3}$ XAS spectra from high-quality single crystals taken from Ref. [33] are very characteristic for whether the cobalt ions are trivalent (nominally d$^6$, in this case $EuCoO_3$ or $Sr_2CoO_3Cl$) or divalent (formally d$^7$, here CoO). There are some subtle differences between the LS and HS variants in the trivalent case, but there are two very prominent low-energy features in the divalent case, highlighted with yellow arrows that make a valence change from tri- to divalent experimentally very easy to spot.

The left-hand panel (a) of Fig. 2 shows the core result of this research. The data from the thin (4 uc) film of LCO (blue) closely resembles a combination of the spectra of the two trivalent model compounds [34]. Upon sandwiching a sample with 2 uc of LCO between LTO layers to generate **two** LTO|LCO interfaces [2*IF] (red), strong XAS intensity can be seen at the location of the two yellow arrows. Now the XAS fingerprint is practically identical to that of CoO: **for this double interface system 100% of the cobalt ions have been transformed into a divalent, $d^7$ electronic configuration**. In the simplest picture this yields a single electron in an otherwise empty $e_g$ orbital manifold. The bottom-most trace in Fig. 2(a) (grey) shows that exactly the same result of complete valence transformation occurs also with the LCO under (mild) compressive strain, as this film stack was grown on bulk LAO.

From the data of Fig. 2 it is already clear that the combination of ultrathin LCO sandwiched between LTO does not support regular trivalent cobalt ions, but rather a divalent state, in line with the theory predictions for charge transfer for this couple in Ref. [17]. **As each of the Co ions in the 2 uc LCO film has picked up an extra electron, the obvious question arising is from where?** The two main possibilities are:
  i) Charge transfer of the $3d^1$ electron from the LTO to the cobalt ions (as takes place in the density functional theory simulations [17])
  ii) or a Co valency change as a result of oxygen (or cation) non-stoichiometry in the LCO layer or anionic/cationic intermixing between the layers.

Obviously, this is an important issue to settle. To work out what is happening in these carefully PLD-grown single- bi- and trilayer systems with layer thicknesses in the few uc level in an unbiased manner, a combination of atomic level structural determination, with sensitive (and non-invasive) valency determination is required. This is precisely what we bring to bear on the problem using soft X-ray XAS, (HA)XPS and TEM techniques.

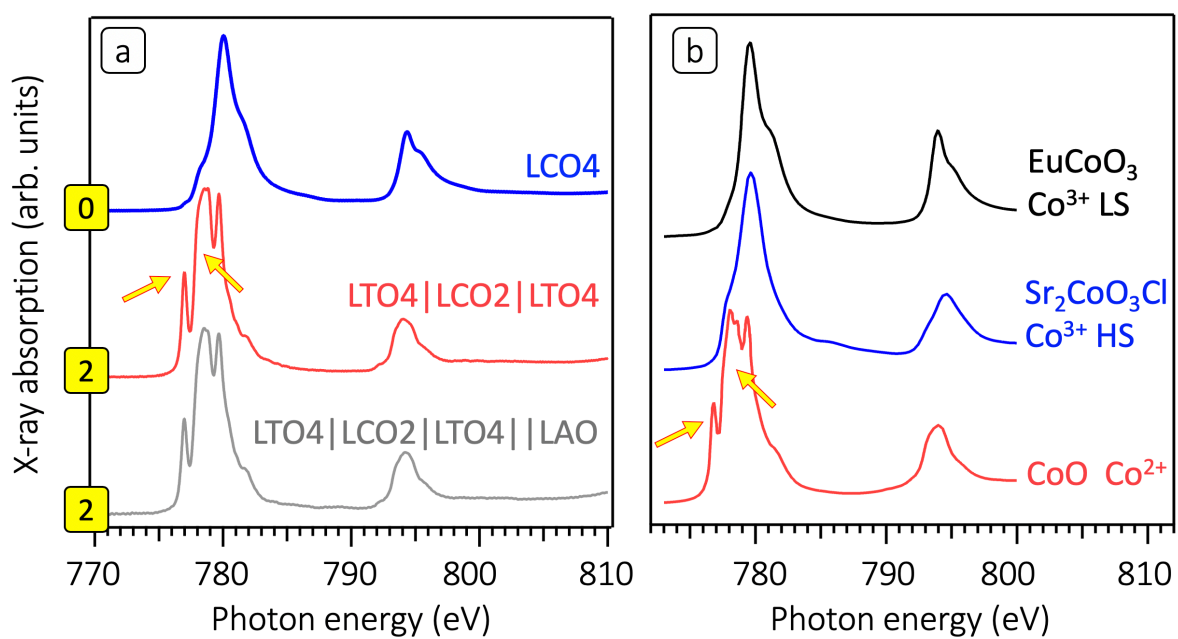

Figure 2: Valency fingerprinting via Co-$L_{2,3}$ XAS shows complete transformation to divalent Co.
(a) From top to bottom: no active interface [0*IF]: 4uc LCO; double IF [2*IF]: 4uc LTO|2uc LCO|4uc LTO; [2*IF] on LaAlO$_3$ substrate: 4uc LaTiO$_3$|2uc LaCoO$_3$|4 uc LaTiO$_3$. Top two datasets measured at 10 K, lowermost at 100 K.
(b) spectra from single crystals of model compounds from Ref. [33]: from top to bottom: EuCoO$_3$ (Co$^{3+}$ low spin) Sr$_2$CoO$_3$Cl Co$^{3+}$ (high spin) and CoO (Co$^{2+}$, high-spin system).

In the following, we discuss data regarding the range within which the cobalt valency is altered in the LCO (from both XAS and STEM); the valency of the Ti in the LTO (from XAS and XPS) and the crystalline quality of the interfacial systems (from STEM). Taken together, these data present a compelling case that charge transfer from LTO to LCO is the leading driver of the phenomena observed. We will gather together all the arguments that bring us to this conclusion again at the close of the paper, once all the different experimental results have been presented[35].

We now move to the **range in the LCO over which this charge transfer effect is active**. This can be probed by varying the LCO thickness in a 4 uc LTO|X uc LCO|4 uc LTO sandwich configuration. The Co-$L_{2,3}$ XAS valency fingerprints for these systems are shown in Fig. 3 for X=2, 4, 6 and 36 uc together with a trace from a single interface, [1*IF] 4 uc LTO |4 uc LCO4 sample for comparison [36].

Starting from the bottom of the stack of spectra in Fig. 3: the [1*IF] data in green already shows the presence of significant divalent cobalt (model compound fits yield 30% divalent Co), linked to it single active LTO|LCO interface.

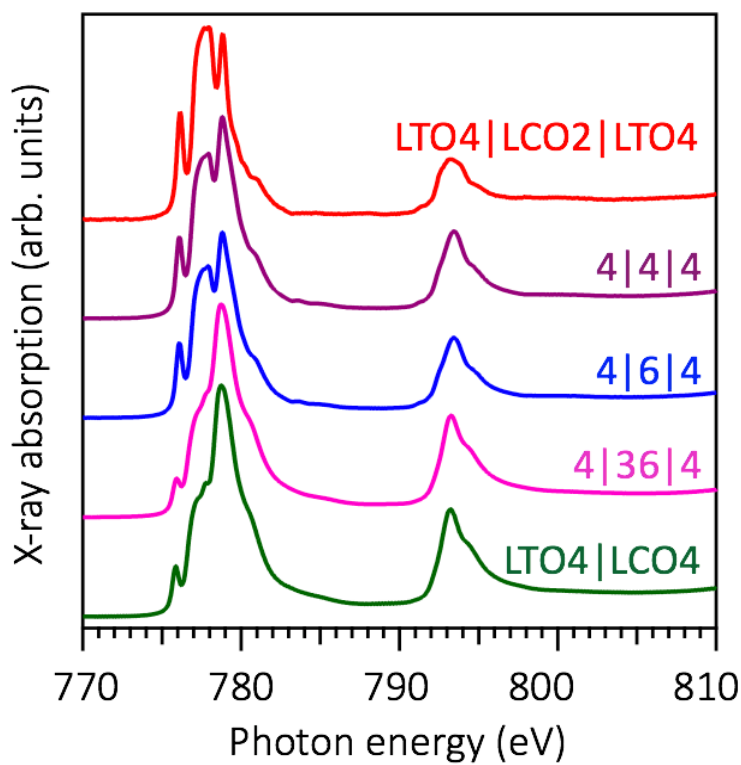

Figure 3: Interfacial character of charge transfer.
Co-$L_{2,3}$ XAS spectra of 2*IF systems: 4uc LTO|X uc LCO|4uc LTO with X=2,4,6, and 36 uc. The bottom-most trace from a [1*IF] 4uc LTO|2uc LCO sample shows the 4|36|4 system is similar

to the single IF case. The LCO thickness dependence **clearly signals the interfacial character of the electron transfer**. All data recorded at 10 K, and individual spectra are offset vertically for clarity.

Next up in Fig. 3 is the [2*IF] system shown in pink. Here the lowermost interface has been placed 15nm below the film surface, and thus is essentially invisible in TEY-XAS due to the latter's 10nm probing depth [27]. The fact that such a [2*IF] sample displays a charge transfer contribution (25% divalent Co) similar to yet slightly smaller than that of the single interface system is a first indication for the **interfacial character of the charge transfer**. The slightly lower charge transfer in the trilayer data stemming from the larger contribution from cobalt ions well away from either active interface than is possible in the thinner [1*IF] sample.

Moving to the centre-most XAS spectrum, confining only 6 uc of LCO between the LTO (blue online) brings both interfaces within measurement range, resulting in 40% divalent Co. As the central LCO layer gets thinner, the divalent percentage increases: in the 4|4|4 sample half the Co ions are divalent, and as discussed in the context of Fig. 2(a), for 4|2|4 all the trivalent cobalt ions have received an additional electron, turning them divalent. Fig. S2 shows how two methods for decomposing the spectra to yield $Co^{2+}$ percentages agree nicely.

The deepest-lying LTO|LCO interfaces for X= 6, 4 and 2 vary from 5.7 to 4.9 to 4.2 nm below the sample surface, so not the 10nm probing depth of TEY-XAS, but a combination of the reduction of the distance in the LCO to the nearest active LTO|LCO interface and a limited range away from the LTO interface at which the charge transfer into the LCO is effective are responsible for the strong growth in divalent character. These data are a second argument for the **interfacial character of the charge transfer process**. The Co-$L_{2,3}$ XAS data tell a clear story: each interfacial contact between two uc's – one LCO, the other of LTO – results in a net transfer of one electron to the LCO system. This is a remarkably large charge transfer. In order to correctly interpret the origin of the electrons picked up by the cobalt ions, atomic-scale information is required on the structure and chemical make-up of the interfacial region.

### Results and discussion – STEM and STEM-EELS data

This brings us naturally to the STEM data. Fig. 4(a) shows a high angle annular dark field (HAADF) STEM image from a cross-section of an LNO-capped LTO4|LCO36|LTO4 film stack grown on STO (substrate not shown). This particular sample was chosen as the upper and lower LTO|LCO interfaces were readily identifiable after FIB-based cross-section preparation. From the image, the high quality of the samples is evident.

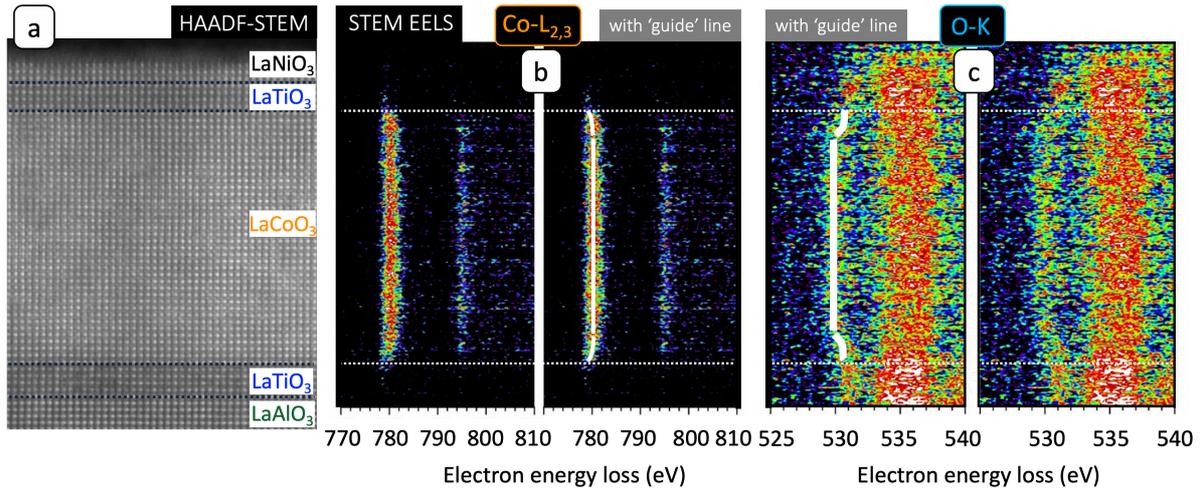

**Figure 4**: TEM imaging and spectroscopy.
(a) HAADF-STEM cross section showing **excellent epitaxy and sharp interfaces**.
(b) & (c) show STEM EELS data from the same cross section. In (b) **a downshift of Co-L$_{2,3}$ main feature close to the active IF**, is consistent with interfacial Co(II). In (c) **an upshift of interfacial O:K pre-peak in LaCoO$_3$** is seen, also consistent with interfacial Co(II). In the central images of (b) and (c) the white lines schematically indicate the energy shifts occurring.
All data recorded at room temperature and in the 2T field of the objective lens.

A 2D representation of the STEM-EELS spatial line-scan data across the two active IFs of the sample are depicted at the Co-L$_3$ and O-K edges in Figs. 4(b) and 4(c), respectively. These 2D spatial/spectroscopic maps, show a definite shifting of the energies of the low energy features of both the Co-L$_3$ edge (downward) and the O:K edge (upward) that take place inside the LCO, within ~3uc of the LTO|LCO interface. A glance at Figs. 2 or 3 suffices to see that the downward shift observed in the Co-L$_{2,3}$ STEM EELS data matches the spectral fingerprint of the divalent cobalt seen in XAS. Importantly, the STEM data give a **direct measure of the range of the charge transfer of ~3 uc** that immediately provides a good qualitative understanding of the behaviour as a function of LCO layer thickness seen in the XAS experiments, such as those shown in Fig. 3.

The second crucial contribution from the TEM data is to rule out prominent cationic non-stoichiometry or migration across the ABO$_3$|AB'O$_3$ interface. Obviously, A-site migration is a non-issue as La has been used throughout all layers in our heterostructures. For the B-site, it would be possible for Ti and Co to migrate across interfaces in both directions and take on different valence states. Both the HAADF-STEM and the EELS data from multiple samples yield no indication - within the error margins of a single unit cell - that such migration occurs, and nor was significant cation non-stoichiometry observed in either material.

The third issue to which the TEM data are very pertinent is that of anionic migration or non-stoichiometry. One can consider the DFT predictions of Ref. [17] as a kind of *gedankenexperiment*: taking perfect 3d$^6$ LCO and perfect 3d$^1$ LTO and putting them in contact yields charge transfer, leading to divalent 3d$^7$ Co & 3d$^0$ tetravalent Ti at and near the interface. In practise, realising this interface requires the sequential growth of two different compounds. On the LCO side of the interface, oxygen vacancies could form to give LaCoO$_{3-\delta}$ with a cobalt valence of 3-(2$\delta$), and on the LTO side, oxygen ion interstitials could exist, forming LaTiO$_{3+\delta}$ with a titanium valence of 3+(2$\delta$). As a matter fact, given the intermediate oxygen pressure that we chose to grow the samples in, one could expect the formation of $LaTiO_{3+\delta}$ and $LaCoO_{3-\delta}$, as this has been seen in the growth of each material as a 'stand-alone' system [23,25].

Thus, in our heterostructures the possibilities are manifold: (a) a $3d^7|3d^0$ $LaCoO_{2.5}$ - $LaTiO_{3.5}$ interface, even (b) a $3d^7|3d^1$ $LaCoO_{2.5}$ - $LaTiO_3$ interface or the reverse: (c) a $3d^6|3d^1$ $LaCoO_3$ - $LaTiO_{3.5}$ interface. Finally, and most remarkably, of course, is the fourth option of a purely electronic charge transfer scenario [17] to yield $3d^7|3d^0$ from the essentially stoichiometric materials. Despite these many options, the experimental evidence for the completed LCO|LTO couple favors the electronic mechanism as we explain in the following.

Firstly, the XAS traces from reference $LaCoO_3$ films grown under identical conditions to the LCO in the bi- and trilayers show divalent Co only at the <3% level. Consequently, although oxygen vacancies in the LCO growth cannot be excluded, they are nowhere near the required level to explain the divalent cobalt signatures in the bi- and trilayers. Secondly, an only minor role for oxygen defects is pointed to by the fact that in STEM we see a clean perovskite structure throughout the whole stack for all types of samples we have presented in this paper. In particular, for $LaTiO_{3+\delta}$ with $Ti^{4+}$, one would expect to see line defects belonging to the 227 - phase [37] as the 113-$ABO_3$ perovskite structure has simply no room for additional large anions. This makes it unlikely the extra oxygen interstitials proposed in Ref. [23] could be present in our heterostructures at a density anywhere near that required to reach $LaTiO_{3.5}$, as the STEM data show such a clean, non-reconstructed $ABO_3$ structure right across the heterointerface.

Next, on the LCO side, also no deviations from the 113-perovskite structure are seen in the TEM imaging. For example, ordered defects are already seen experimentally for 10% oxygen vacancies, equivalent to a composition of $LaCoO_{2.7}$ [38]. Our TEM data are wholly devoid of the stripe- or line-like features that have been interpreted as Co spin-state ordering or O-vacancy ordering in LCO [39-43]. In epitaxial films one would also expect a measurable expansion of the LCO c lattice parameter [44] something which is not seen in our STEM data. To close-up the discussion at this stage of the presented experimental data, we also mention that the fact that the divalent Co is only formed close to the interface with LTO is also naturally explained in the charge transfer scenario, but the special spatial arrangement of possible oxygen vacancies required to achieve this lacks credible motivation. We will return again to the key question of electronic charge transfer vs. anion migration or non-stoichiometry at the end of the paper, after all the experimental data have been presented and interpreted.

### Results and discussion – HAXPES & XPS data

Keeping with the electronic configuration of the cobalt ions for a little longer, Fig. 5 shows Co2p HAXPES spectra of a [0*IF] 4 uc LCO sample (Nb:STO substrate), [1*IF] 4 uc LTO|4 uc LCO and [2*IF] 4 uc LTO|4 uc LCO|4 uc LTO, the latter grown on LAO substrates, together with reference spectra from bulk LCO and CoO from the literature [45,46]. The Co 2p photoemission lineshape is a result of numerous final-state charge-transfer and multiplet interactions and this cautions against a strict quantitative analysis. However, the absence or presence of characteristic satellite structures appearing at ~6eV higher binding energy than the main spin-orbit-split main lines are a tell-tale sign of the Co valence. These features are indicated in Fig. 5 using yellow arrows, and are a result of a so-called 'shake-up' process in which ligand-to-metal charge transfer takes place, yielding a $|2p^5 3d^{n+1}\underline{L}\rangle$ final state [47]. In octahedrally coordinated Co[III] systems such as bulk LCO, the charge transfer energy, $\Delta$, required to do this is too great and no satellite is observed [48], but in divalent Co[II] this is a salient feature of the spectrum [49].

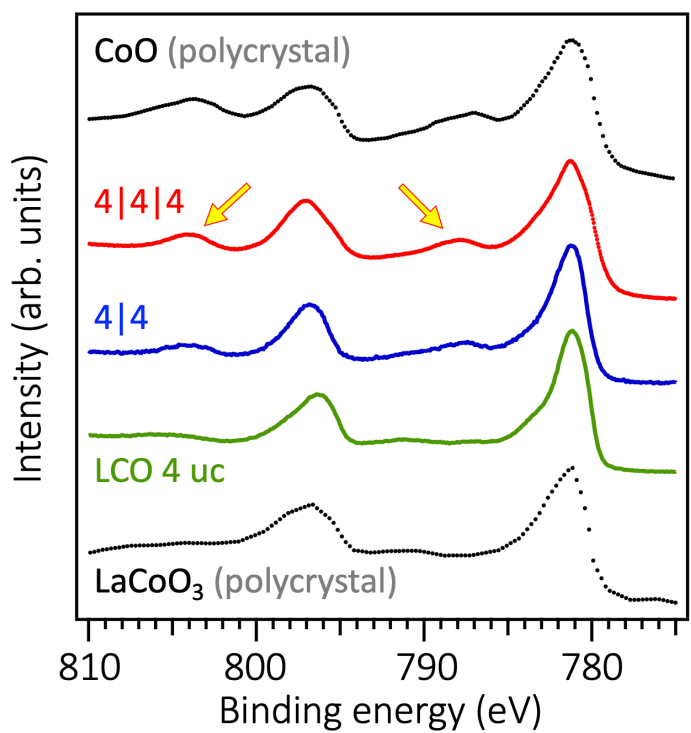

**Figure 5: XPS confirms Co(III) $d^7$ ions in interface samples:** HAXPES spectra of the Co2p core level of (top to bottom): CoO [Ref. 46], [2*IF] LTO4|LCO4|LTO4 and [1*IF] LTO4|LCO4 on LAO substrates and LCO 4 uc on a Nb:STO substrate (online red, blue green, respectively); bulk LCO [Ref. 45].
The thin film data were recorded using hν=2.2 keV at 250 K, and the yellow arrows highlight the shake-up satellites characteristic of octahedrally coordinated Co(II) ions. Individual spectra are offset vertically for clarity.

Looking at the data of Fig. 5, it is clear that the [1*IF] 4|4 sample shows clear satellite intensity (and thus presence of Co[II] ions) and that this increases further in the case of the [2*IF] 4|4|4 sample, **indicating the interfacial character of the charge transfer to LCO**, in excellent agreement with the XAS and TEM-EELS data. In line with this, for the 4 uc LCO film without an 'active' LTO|LCO interface, no shake-up satellite is observable, just as for the bulk LCO spectrum from trivalent cobalt. Given the thinness of the cobaltate layers and the significant inelastic mean-free pathlength of the photoelectrons under these conditions, no depth-profiles could be extracted from the angle-dependent HAXPES data.

Core level photoemission and X-ray absorption can also be used to examine the valence state and associated d-electron count of the Ti in the LTO. In Fig. S3, the Ti-$L_{2,3}$ edges of the LTO layers in [2*IF] samples comprised of 2, 4 and 6 uc of LCO between a pair of 4 uc LTO layers are shown, as well as that of STO. The intensity ratio of the first two ($L_3$ $t_{2g}$- and $e_g$-related) structures are different in the trilayers compared to STO, but otherwise - at a glance - all four traces clearly signal Ti in the tetravalent ($3d^0$) state. On a qualitative level this can argue for the transfer of the 3d electron of the LTO to LCO, as predicted in [17]. However, the Ti-$L_{2,3}$ XAS spectra of Fig. S3 do signal essentially 100% $Ti^{4+}$ in all three of the 4|6|4, 4|4|4 and 4|2|4 stacks, yet comparing to the Co valence fingerprints in Fig. 3, the charge book-keeping does not balance up perfectly: it appears that more Ti $3d^1$ electrons have transferred than there are Co ions to house them. We do not have a clear explanation for this imbalance.

The question of $Ti^{4+}$ in LTO has been addressed in part in Ref. [23], although the extra oxygen interstitials they propose would seem difficult to fit into 113-$ABO_3$ structure at the required density. In Fig. S4, we present in-situ Ti2p XPS data indicating the situation in the reality of the thin film growth process is quite subtle. Here an LTO film is grown on LAO under the conditions given in Table I, and then transferred without breaking vacuum from the PLD to the XPS system. After post-growth cool down, the Ti2p spectrum shows tetravalent Ti, rather than trivalent state nominally expected. Subsequent annealing of the LTO film in the UHV environment of the XPS system boosts the trivalent Ti core level feature, and after re-cooling from 700°C, the top-most spectrum in Fig. S4(a) shows that the $Ti^{3+}$ ($3d^1$) configuration belonging to stoichiometric LTO is dominant. Fig. S4(b) shows that the $Ti^{3+}$ feature can be cycled 'on' (after cooling) or 'off' (at elevated temperature). Ti vacancies can be excluded as the main driver from these data. These in-situ XPS experiments – in line with those of Ref. [23] - show that $Ti^{3+}$ in LTO films can certainly be generated, but in the context of this paper one should add under conditions that are too aggressively reducing for LCO to remain stable.

In addition, to the true electron count in the LTO structures, it is also conceivable that Ni3d level in the LNO cap is positioned in energy such that it can act as an acceptor for electrons from the LTO, or indeed from the LCO. LNO was not covered as an example in Ref. [17], and perhaps both the electron counting imbalance mentioned above, and the smaller divalent Co contribution seen in thin, [1*IF] samples relative to the expectation from the 3 uc charge transfer range from STEM have a connection to the LNO. However, a detailed study of the role of LNO or other capping materials is beyond the remit of this paper.

Results and discussion – testing the range of charge transfer

The penultimate results section of the paper presents a double-check of the range over which the charge transfer between LTO and LCO is active. The calculations of Ref. [17] also suggested the possibility of modulation doping, in which there is spatial separation between the location of the (potentially) conduction electrons and the dopants/structures that give rise to the charge transfer. Up to 5 uc of a transition metal oxide $SrZrO_3$ buffer layer (what we call below a 'break' layer) is suggested to be able to leave the interfacial charge transfer unaffected [17]. Given the very robust dependability of the PLD growth, growth of samples including such 'break' layers provides a simple, combined test of both the interfacial nature of the charge transfer and the practical feasibility of modulation doping in such oxide heterostructures.

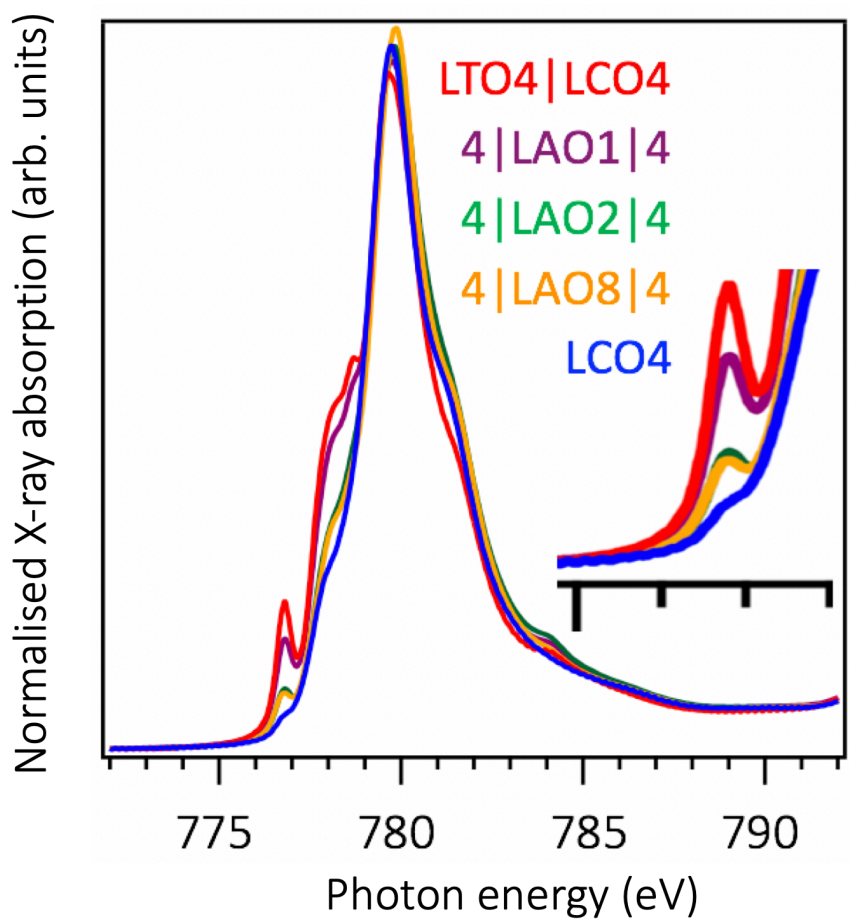

Figure 6: Testing 'modulation doping' style charge transfer idea using LaAlO$_3$ 'break' layers: Co-L$_3$ XAS traces of 4 uc LTO|x uc LAO|4 uc LCO, with x=0,1,2,8. LAO interrupts the charge transfer, approaching the zero-active-IF 4uc LCO system. The inset shows a zoom of the pre-peak region.

In Fig. 6 the main Co-L$_3$ XAS feature is shown for a series of [1*IF] samples. The LTO4|LCO4 (red online) and LCO4 (blue online) have been discussed before, with the former showing 25% divalent Co and the latter <3%. The three traces in between possess a single IF, but with a varying number (here 1, 2 and 8) unit cells of LAO slipped in between as an electronic 'break' layer (see Fig. 1[d]). As LaAlO$_3$ is a wide band gap insulator, with no variable oxidation state cations, it is a highly effective charge transfer circuit breaker. The inset highlights the lowest energy divalent Co pre-peak feature, and clearly shows what is happening: charge transfer at a level comparable to that in the LTO4|LCO4 system is maintained despite addition of a single uc of LAO as a 'break'. However, already only two uc LAO break reduces the tell-tale divalent Co feature by a factor 2.5.

This simple test ties in nicely with the information from the XAS and STEM EELS data of the trilayers: that the **charge transfer process is interfacial**, and it has a range of order 2-3 uc. Thus, the DFT-based expectation of interfacial charge transfer itself, and its ability to bridge a thin insulating barrier from the O2p band alignment idea coming from the DFT work holds up in these experiments on high quality thin films heterostructures.

Results and discussion – XMCD data & sum rule analysis

In the final results part of the paper, we turn to the spin state and magnetic properties of the Co ions in the LCO films. Obviously, there is no straightforward manner in which the individual

Co[II] ions with Co $3d^7$ configuration can be spinless. Measuring the magnetic properties of complex oxides in the form of ultrathin films as components in heterostructures on bulk substrates is a severe technical challenge for regular magnetometry. In such cases, the chemical specificity offered by XMCD [50,51] offers not only exquisite sensitivity, but also – via application of XMCD sum rules [52] - extraction of both the spin and orbital moments.

Panels (a-d) Fig. 7 show XMCD data recorded in different applied magnetic fields for **0, 1** and **2\*IF** systems. Figure S5 shows exemplary raw $\sigma^+$ and $\sigma^-$ XAS data that yield the XMCD signal. Each XMCD panel in Fig. 7 is on the same y-scale in units of the % of the maximal Co-$L_3$ absorption. As is expected for a thin film system [21], the **0\*IF** 4 uc LCO sample is not perfectly low spin, with the data of Fig. 7 showing an XMCD signal at the 7-8 percent level at the maximal field of 14T. Adding an active LTO|LCO interface (**1\*IF**: LTO4|LCO4) alters the spectral form of the XMCD signal and it more than doubles in magnitude. On going to the **[2\*IF]** systems LTO4|LCO4|LTO4 and LTO4|LCO2|LTO4, the XMCD signal is four- and six-fold enhanced compared to that of the single LCO layer. Thus, it is very clear that **the controlled charge transfer between LTO and LCO is turning the interfacial cobalt ions into magnetically-polarisable entities**, in keeping with the odd number of electrons in their d-shell from the XAS lineshape analysis.

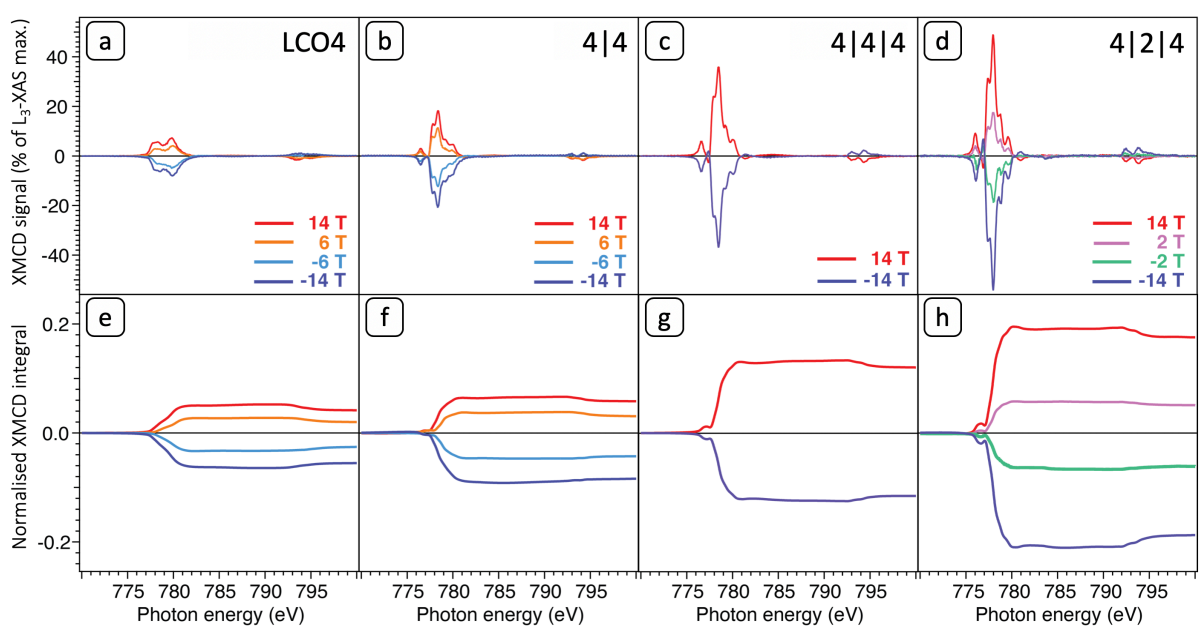

**Figure 7**: XMCD shows boosted Co magnetic moment on formation of Co(III) $d^7$ ions in interface samples: (a-d) TEY XMCD signal ($\sigma^+$-$\sigma^-$) at the Co-$L_{2,3}$ edges as a percentage of the maximum of the Co-$L_3$ absorption. Panels (e-h) show the integrals over the XMCD signals, normalized to the integral of the unpolarized absorption for the same data. (a,b) are for **0\*IF**: 4 uc LCO; (c,d) **1\*IF**: 4uc LTO|4uc LCO; (e,f) are **2\*IF**: 4uc LTO|4uc LCO|4uc LTO and (g,h) are **2\*IF**: 4uc LTO|2uc LCO|4uc LTO. All data are measured at 10K with the samples cooled in the fields shown.

The XMCD sum-rules enable extraction of more quantitative information on the magnetic properties [52]. These involve the integration of the XMCD signal over the photon energy region of the Co-$L_2$ and $L_3$ edges, and panels (e-h) of Fig. 7 show these normalised XMCD integrals. Already without any further analysis, the fact that the XMCD integral shows a downward step at the L$_2$ edge and maintains a finite value thereafter indicates that the Co ions possess both spin *and* orbital moments.

The spin and orbital moments extracted from the sum rule analysis are shown in Table II. Increasing the number of active interfaces boosts both the spin and orbital moments by factors equal to or exceeding 1.2 [**1*IF**], 1.5 [**2*IF** with LCO4] and 2.4 [**2*IF** with LCO2]) compared to **0*IF** (LCO4). **For all samples, the orbital moment is considerable, amounting to half of the spin moment, and is aligned parallel to the spin moment**, meaning the g-factor exceeds 2 in these systems.

Grown under coherent epitaxial strain on bulk STO, the LCO layers in these systems are under in-plane tensile strain: c/a<1. Studies of 10nm thick films of CoO grown either sandwiched between MnO on a Ag substrate or grown directly on silver [53] have shown that the 3d spin-orbit interaction prevents the collapse of the orbital moment [54] for high spin Co 3d$^7$ states when c/a<1. We propose that this same mechanism is likely to be operative in the in-plane tensile strained LCO films presented here.

The data presented in Fig. 7 and their analysis clearly point to the generation of spinful Co3d$^7$ entities as soon as one or more active LTO|LCO interfaces are introduced, with a total moment of 1.16 μ$_B$, and **m$_l$/m$_s$** of 0.5 for the fully divalent cobalt ions in LTO4|LCO2|LTO4.

Table II. Magnetic parameters from XMCD sum rule analysis for -14T applied field at 10K.

| System | % Co[II] from XAS | Orbital moment, m$_l$ (μ$_B$) from XMCD | Spin moment, m$_s$ (μ$_B$) from XMCD | m$_l$ / m$_s$ | Total moment, m$_{tot}$ (μ$_B$) from XMCD | Saturation magnetisation M$_{sat}$(μ$_B$) from Brillouin function | Factor increase in m$_{tot}$ cf. LCO4 |
|---|---|---|---|---|---|---|---|
| **0*IF** LCO4 | <3 | 0.15 | 0.33 | 0.45 | 0.48 | 0.53 | - |
| **1*IF** LTO4|LCO4 | 25 | 0.2 | 0.38 | 0.55 | 0.58 | 0.64 | 1.2 |
| **2*IF** LTO4|LCO6|LTO4 | 40 | 0.26 | 0.5 | 0.52 | 0.76 | 0.89 | 1.2 |
| **2*IF** LTO4|LCO4|LTO4 | 50 | 0.26 | 0.47 | 0.54 | 0.73 | 0.81 | 1.5 |
| **2*IF** LTO4|LCO2|LTO4 | 100 | 0.37 | 0.75 | 0.50 | 1.16 | 1.29 | 2.4 |

In order to discuss the magnetic behaviour of the Co ions, Fig. 8 presents the field and temperature dependence of the magnetisation extracted from the XMCD sum rules for a **0*IF** LCO sample and two **2*IF** systems: LTO4|LCO6|LTO4 and LTO4|LCO2|LTO4. The colour-coded solid lines are fits to: $M = M_{sat} B_J(x)$, where the experimentally determined **m$_l$/m$_s$** ratio and the S expected from the valence observed in the XAS were used to calculate J using a Brillouin function

$$B_J(x) = \frac{2J+1}{2J} coth\left(\frac{(2J+1)x}{2J}\right) - \frac{1}{2J} coth\left(\frac{x}{2J}\right) \text{ with } x = \frac{g_J \mu_B J B}{k_B T}.$$

The Landé g-value is

$$g_J = \frac{3}{2} + \frac{S(S+1) - L(L+1)}{2J(J+1)},$$

and we note that such a Brillouin function describes the spin physics of a collection of spins behaving paramagnetically [55].

The saturation magnetisation, **M$_{sat}$**, was varied to achieve optimal fits to the experimental field- (main figure) or T-dependence (inset) of the data. **M$_{sat}$** is about 10% greater than the values connected to maximal[minimal] fields[temperatures] accessed experimentally. Simple inspection of Fig. 8 shows the Brillouin function captures the essence of the field dependence and the strong decay of the magnetisation as temperature is raised for all the samples measured, indicating **paramagnetic behaviour**.

For the **0*IF** system LCO4, the 3d$^6$ HS (S=2) population provides the paramagnetic magnetic response [56], and taking into account the experimentally determined $m_l/m_s$ ratio of 0.45, a good fit is achieved for a saturation magnetisation of **$M_{sat}$**(LCO4) = 0.45 $\mu_B$ per Co atom.
For the **2*IF** systems LTO4|LCO6|LTO4 and LTO4|LCO2|LTO4, in which charge transfer has created divalent Co 3d$^7$ ions, there are two options
- a high spin $t_{2g}^5 e_g^2$ S=3/2 state (HS) or
- a low spin $t_{2g}^6 e_g^1$ S=1/2 state (LS).

The Brillouin function fits in the main panel of Fig. 8 are for HS 3d$^7$, the same spin state as in CoO (the latter is also HS as a thin film under either compressive or tensile in-plane strain [53]). Without exception, divalent cobalt oxides are quoted as HS in the literature, as their on-site Coulomb interaction energy, $U_{dd}$ (at the root of Hund's first rule) is greater that the crystal field energy. In the mixed valent cobalt oxide La$_{1.5}$Sr$_{0.5}$CoO$_4$, successful modelling of the Co-L$_{2,3}$ XAS using a combination of the spectra from EuCoO$_3$ (LS Co3d$^6$) and CoO (HS Co3d$^7$) is provided as evidence that the divalent Co in La$_{1.5}$Sr$_{0.5}$CoO$_4$ is HS [57]. The fact that the XAS data from our bi- and trilayer samples can also be fitted in exactly the same way (see Fig. S2) argues by analogy for the conventional HS spin state for the divalent Co 3d$^7$ here. Having said that, the Brillouin function fit for the field dependence for the LS 3d$^7$ state, only gives a marginally worse fit. The saturation magnetisations - using $m_l/m_s$ ratios of 0.52 for both [2*IF] samples – are 0.89 and 1.29 $\mu_B$ per Co atom for LTO4|LCO6|LTO4 and LTO4|LCO2|LTO4, respectively, as shown in Table II.

The expected spin-only magnetic moment in a simple single-ion picture for HS Co 3d$^7$ would be 3$\mu_B$ per Co, and under similar assumptions would be 1$\mu_B$ for LS Co 3d$^7$. In this straightforward view, the LS configuration can be argued to yield a spin moment closer to the experimental spin-only value of 0.75$\mu_B$ for the fully Co 3d$^7$ LTO4|LCO2|LTO4 system. Arguments can also be made that the improved Goldschmidt tolerance factor for the smaller LS Co3d$^7$ ion [58] of (t=0.958) compared to the HS one (t=0.915) could also help counteract the additional Coulomb repulsion cost of the LS state.

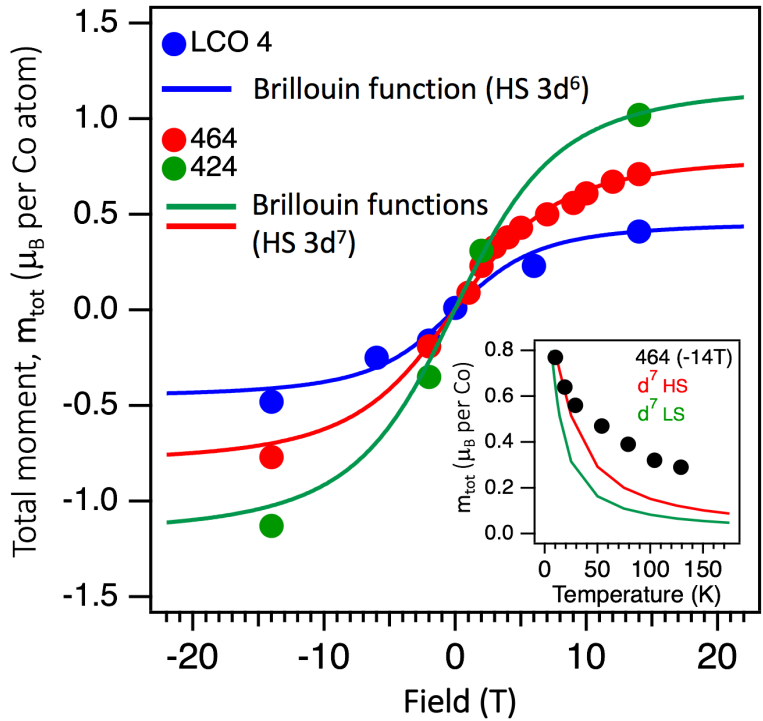

Figure 8: **Magnetic field and temperature dependence of the cobalt element-specific magnetisation shows Brillouin-function behaviour.** Shown is the field dependence of $m_{tot}$, the sum of spin and orbital moments (coloured symbols), determined using the XMCD sum rules for **0*IF**: 4 uc LCO (blue), and **2*IF**: 4uc LTO|6uc LCO|4uc LTO (red) and 4uc LTO|2uc LCO|4uc LTO (green). The solid lines show fits using a Brillouin function ($T_{sample}$ 10K) for S values from the XAS valence analysis (S=3/2 for HS Co $3d^7$; S=2 for HS Co $3d^6$) and L values matching the determined $m_l/m_s$ values from the sum rules. The inset shows the temperature dependence of the sum of spin and orbital moments for the **2*IF** system 4uc LTO|6uc LCO|4uc LTO (black symbols) measured in a field of -14 T. The lines show the Brillouin function behaviour expected for HS Co $3d^7$ HS (red) and LS Co $3d^7$ (green).

The inset to Fig. 8 shows the temperature dependence of total magnetic moment of the **2*IF** system LTO4|LCO6|LTO4 in the range of 10 to 180 K. Example XMCD spectra underlying the data-points of Fig. 8 are shown in Fig. S6(a). The solid lines are from the Brillouin functions (red: HS $3d^7$, S=3/2, $m_l/m_s$=0.53; green line: $3d^7$ LS, S=1/2, $m_l/m_s$=0.53) with $M_{sat}$ = 0.89$\mu_B$. The high spin curve yields a better result than for the low spin state. Both fit curves show steeper decay of the magnetisation than do the data, and suggest a contribution to the experimental value of the total moment from another source that grows as temperature is raised. A natural candidate for this is a trivalent Co $3d^6$ HS contribution (S=2) growing from zero at 10K to of order 15% at 150K [21]. Seeing as 60% of the Co in the LTO4|LCO6|LTO4 system for which we have the detailed field- and temperature-dependent XMCD data is simply trivalent LCO, it is not unreasonable to suggest that this is the source of the additional magnetisation at higher temperatures.

In any case, the lack of remnant magnetisation in zero-field, and the good description provided by the Brillouin function at low temperatures argues firmly against ferromagnetism in the case of these bi- and tri-layer interfacial Co spin systems[59]. In keeping with this, supplementary Fig. S6(b) shows a lack of in-plane/out-of-plane anisotropy in the XMCD of the **2*IF** system LTO4|LCO4|LTO4. It is also clear from Fig. 8 that these interfacial systems are not generating

a long range ordered AFM ground state. AFM Co-Co correlations, however, have been cited as the cause of reduced moments for the paramagnetic, divalent Co sites in FM films of Co-doped ZnO studied using XMCD [60], and in thick, Ce-doped LCO films (also containing HS Co $3d^7$), reduced moments are also reported [61]. Thus, AFM correlations or local patches of AFM order could be responsible for the reduced saturation magnetisation of the Co $3d^7$ ions from the XMCD analysis on these bi- and trilayers. Detailed angular and T-dependent XMLD measurements would be required to examine possible (incipient) AFM ordering while distinguishing these magnetic contributions from charge order/anisotropies in these non-cubic systems [62]. Such experiments go well beyond the remit of this investigation.

The bottom line of the XMCD experiments on the magnetic properties is that the **interfacial charge transfer clearly increases the spin and orbital moments, with the Co $3d^7$ spins displaying paramagnetic behaviour**.

The rich spin physics of cobalt oxides also sheds light on the lack of electrical conduction in these systems, also for cases with non-integer 3d electron counts. If the divalent Co $3d^7$ are in a high spin state and neighbouring trivalent Co $3d^6$ are low spin, then a process called spin-blockade prohibits hopping, suppressing conduction, as has been suggested in $HoBaCo_2O_{5.5}$ [63], $La_{1.5}Sr_{0.5}CoO_4$ [57] and Ce-doped LCO films [61]. For the $3d^7$ integer electron count in the **2*IF: 4uc LTO|2uc LCO|4uc LTO** system, Mott physics most likely lies at the root of the insulating ground state.

Before concluding, we return to the key discussion of the microscopic origin of the remarkable charge transfer to create structurally unaltered, paramagnetically polarisable $Co^{2+}$ at a 100% level across all LCO unit cells abutting LTO in high quality, non-polar oxide heterointerfaces. The final result, that the Mott insulator LTO has lost its d-electron and that the charge transfer insulator LCO has picked up an extra electron is unequivocal: **the prediction that the LCO|LTO couple exists as $3d^7|3d^0$ and not $3d^6|3d^1$ is proven correct.**

Either electronic charge transfer or oxygen non-stoichiometry or migration are the two front-runner mechanisms for the observed behaviour. Here we sum up all the arguments favouring the electronic mechanism as the dominant one.

- Although oxygen vacancies will surely exist, the single component LCO films grown in an identical manner as the heterointerface systems show them to be present only at the few % level – never enough to get full valence-switching as seen in the trilayers. In the heterostructure 'break' samples, such as that with 8 uc of LAO, a ceiling of 5% also follows from the Co XAS lineshape.
- The STEM data are clear in their exclusion of oxygen vacancy ordering at any sufficient density in the LCO, nor do they allow structural alterations such as line defects associated with domains of $La_2Ti_2O_7$ in the LTO.
- Under PLD growth conditions, oxygen vacancies are mobile, and thus not only would additional arguments be required to explain them being pinned at the interfaces, but also a thin break layer of a few uc of $LaAlO_3$ would not impede vacancy diffusion into LCO, whereas it *can* provide an isolation layer, blocking interfacial charge transfer.

Given the ease with which excess oxygen or oxygen vacancies are created in late transition metal oxides, it is interesting to consider why they are not the dominating microscopic mechanism for the charge transfer here. Evidently, although the differences in 3d energy levels that ultimately drives the electronic charge transfer could lead to a more chemical expression of energy lowering, we speculate that upon fabrication at high temperature, having LTO and LCO side-by-side presents a different thermodynamic situation than having single layers of LTO or LCO on their own. As the heterointerface is grown, the LTO already has the possibility of offloading its 'unwanted' 3d electron to the LCO, with no necessity for the formation of either oxygen interstitials or vacancies.

## Conclusions and outlook

We have grown high-quality thin films of the charge transfer insulator LCO using PLD possessing either 0, 1 or 2 structurally abrupt, non-polar interfaces to the Mott insulator LTO. The data clearly show the **O2p-band-alignment-based DFT prediction of interfacial charge transfer to yield a Co3d$^7$ | Ti 3d$^0$ electronic configuration is correct**. The heterointerface, although non-conducting due to spin blockade physics, provides a tunable population of interfacial, most-likely high-spin, divalent Co.

The magnitude of the electron transfer for the bi- and trilayers is one electron per interfacial LTO|LCO unit cell, per interface in the structure. Therefore, the LCO film thickness and the number of LTO|LCO interfaces provide a pair of deterministic control knobs for the charge transfer. Additionally, the charge transfer can be reduced via thickness of a 'break' layer between the LTO and LCO. The interfacial 3d$^7$ Co ions formed by charge transfer exhibit significant orbital moment, likely due to a combination of the anisotropic crystal field and Co 3d spin-orbit coupling, and detailed XMCD investigations point clearly to their paramagnetic behaviour.

These experiments clearly affirm the **O 2p band alignment concept as a new and successful design philosophy for the engineering of strongly correlated quantum materials** [17], without the need for the system to be responding to an incipient polar catastrophe, and without introducing non-stoichiometry or the cationic disorder connected to chemical doping. To be able to do this is of interest in controlling conductivity, magnetic states and also (catalytic) chemical reactivity.

## Acknowledgements

This research is part of the NWO/FOM research programme DESCO (VP149), which is financed by the Netherlands Organisation for Scientific Research (NWO). D.K., J.V. and N.G. acknowledge funding from the *Geconcentreerde Onderzoekacties* (GOA) project "Solarpaint" of the University of Antwerp. The Qu-Ant-EM microscope used in this study was partly funded by the Hercules fund from the Flemish Government.
The authors are grateful to David McCue and Mark Sussmuth and Paul Steadman at Diamond Light Source for excellent user support. The synchrotron-based experiments have been supported by the project CALIPSOplus under the Grant Agreement 730872 from the EU Framework Programme for Research and Innovation HORIZON 2020. SKM thanks the Department of Science and Technology, India (SR/NM/Z-07/2015) for the financial support and Jawaharlal Nehru Centre for Advanced Scientific Research (JNCASR) for managing the project.

## Author contributions

M.S.G., G.A.K. J.G. and G.K. conceived and planned the XAS and HAXPES experiments. J.G., G.K. and G.R. synthesized the samples and performed lab XPS and data analysis. M.S.G., G.A.K. J.G. S.K.M., S.S., X.V. carried out the synchrotron work, with the data analysis done by G.A.K. and M.S.G., while P.B., T.-L.L., and C.S. provided essential user support at the synchrotron. N.G. D.K. and J.V. carried out TEM-EELS experiments and their analysis. M.S.G. and G.A.K. wrote the manuscript and all authors contributed to the discussion and revision of the manuscript.

# References

[1] K. Mizushima, P.C. Jones, P.J. Wiseman, J.B. Goodenough Mat. Res. Bull. Vol **15** 783-789 (1980)

[2] Foster, F S, C J Pavlin, K A Harasiewicz, D A Christopher, and D H Turnbull. "Advances in Ultrasound Biomicroscopy" Ultrasound in Medicine & Biology **26**, 1 (2000).

[3] Shrout, Thomas R., and Shujun J. Zhang. "Lead-Free Piezoelectric Ceramics: Alternatives for PZT?" Journal of Electroceramics **19** 111 (2007)

[4] M. Imada, A. Fujimori and Y. Tokura, "Metal Insulator Transitions", Rev. Mod. Phys. **70** 1039 (1998)

[5] Y. Tokura and N. Nagaosa, "Orbital Physics in Transition-Metal Oxides", Science **288**, 462 (2000),

[6] A. Ohtomo and H. Y. Hwang, "A High-Mobility Electron Gas at the $LaAlO_3/SrTiO_3$ Heterointerface." Nature **427**, 423 (2004)

[7] Brinkman, A, M Huijben, M Van Zalk, J Huijben, U Zeitler, J C Maan, W G Van Der Wiel, G Rijnders, D H A Blank, and H Hilgenkamp, "Magnetic Effects at the Interface between Non-Magnetic Oxides", Nature Materials **6**, 493 (2007)

[8] N. Reyren, S. Thiel, A. D. Caviglia, L. F. Kourkoutis, G. Hammerl, C. Richter, C. W. Schneider, T. Kopp, A.-S. Ruetschi, D. Jaccard, M. Gabay, D.A. Muller, J.-M. Triscone, J. Mannhart, "Superconducting Interfaces Between Insulating Oxides", Science **317**, 1196 (2007)

[9] Caviglia, A D, S Gariglio, N Reyren, D Jaccard, T Schneider, M Gabay, S Thiel, G Hammerl, J Mannhart, and J-M Triscone, "Electric Field Control of the $LaAlO_3/SrTiO_3$ Interface Ground State", Nature **456**, 624 (2008)

[10] Nakagawa, Naoyuki, Harold Y. Hwang, and David A. Muller, "Why Some Interfaces Cannot Be Sharp", Nature Materials **5**, 204 (2006)

[11] Kalabukhov, Alexey, Robert Gunnarsson, Johan Börjesson, Eva Olsson, Tord Claeson, and Dag Winkler. "Effect of Oxygen Vacancies in the $SrTiO_3$ Substrate on the Electrical Properties of the $LaAlO_3$ $SrTiO_3$ Interface." Physical Review **B75**, 2 (2007)

[12] Chambers, S A, L Qiao, T C Droubay, T C Kaspar, B W Arey, and P V Sushko. "Band Alignment, Built-In Potential, and the Absence of Conductivity at the $LaCrO_3/SrTiO_3(001)$ Heterojunction." PHYSICAL REVIEW LETTERS **107**, 206802 (2011)

[13] Slooten, E, Zhicheng Zhong, H J A Molegraaf, P D Eerkes, S de Jong, F Massee, E van Heumen, Kruize, M K Wenderich, S, Kleibeuker, J E, Gorgoi, M, Hilgenkamp, H, Brinkman, A, Huijben, M, Rijnders, G, Blank, D H A Koster, G, Kelly, P J, and Golden, M S, "Hard X-Ray Photoemission and Density Functional Theory Study of the Internal Electric Field in $SrTiO_3/LaAlO_3$ Oxide Heterostructures." Physical Review **B87** 85128 (2013)

[14] Liao, Z., M. Huijben, Z. Zhong, N. Gauquelin, S. Macke, R. J. Green, S. Van Aert, J. Verbeeck, G. Van Tendeloo, K. Held, G. A. Sawatzky, G. Koster & G. Rijnders, "Controlled Lateral Anisotropy in Correlated Manganite Heterostructures by Interface-Engineered Oxygen Octahedral Coupling." Nature Materials **15**, 425 (2016)

[15] Pavlo Zubko, Stefano Gariglio, Marc Gabay, Philippe Ghosez, Jean-Marc Triscone, "Interface Physics in Complex Oxide Heterostructures", Annual Review of Condensed Matter Physics, **2**, 141 (2011)

[16] Herbert Kroemer, "Nobel lecture: Quasielectric Fields and Band Offsets: Teaching Electrons New Tricks", Reviews of Modern Physics **73**, 783 (2001)

[17] Zhong, Zhicheng, and Philipp Hansmann. "Band Alignment and Charge Transfer in Complex Oxide Interfaces." Physical Review X **7** 011023 (2017)

[18] Kleibeuker, J. E., Z. Zhong, H. Nishikawa, J. Gabel, A. Müller, F. Pfaff, M. Sing, Held, K. Claessen, R. Koster, G. Rijnders, "Electronic Reconstruction at the Isopolar $LaTiO_3/LaFeO_3$ Interface: An X-Ray Photoemission and Density-Functional Theory Study." Physical Review Letters **113** 237402 (2014)

[19] J. Suntivich, K. J. May, H.A. Gasteiger, J.B. Goodenough and Y. Shao-Horn, Science **334**, 1383 (2011)

[20] Saitoh, T., T. Mizokawa, A. Fujimori, and M. Abbate. "Electronic Structure and Temperature-Induced Paramagnetism in $LaCoO_3$." Physical Review **B55** 4257 (1997)

[21] Haverkort, M. W., Hu, Z., Cezar, J. C., Burnus, T., Hartmann, H., Reuther, M., Zobel, C., Lorenz, T., Tanaka, A., Brookes, N. B., Hsieh, H. H., Lin, H.-J., Chen, C. T. & Tjeng, L. H. "Spin State Transition in LaCoO$_3$ Studied Using Soft X-ray Absorption Spectroscopy and Magnetic Circular Dichroism". Phys. Rev. Lett., **97**, 176405 (2006)

[22] Gertjan Koster, Guus Rijnders, Dave H.A. Blank, Horst Rogalla "Surface morphology determined by (001) single Crystal SrTiO$_3$ termination" Physica C **339**, 215 (2000)

[23] Philipp Scheiderer, Matthias Schmitt, Judith Gabel, Michael Zapf, Martin Stübinger, Philipp Schütz, Lenart Dudy, Christoph Schlueter, Tien-Lin Lee, Michael Sing, and Ralph Claessen "Tailoring Materials for Mottronics: Excess Oxygen Doping of a Prototypical Mott Insulator" Adv. Matter. **30** 1706708 (2018)

[24] A. Ohtomo, D.A. Muller, J.L. Grazul and H.Y.Hwang "Epitaxial growth and electronic structure of LaTiO$_x$ films" Appl. Phys. Lett. **80**, 3922 (2002)

[25] Dechao Meng, Hongli Guo, Zhangzhang Cui, Chao Ma, Jin Zhao, Jiangbo Lu, Hui Xu, Zhicheng Wang, Xiang Hu, Zhengping Fu, Ranran Peng, Jinghua Guo, Xiaofang Zhai, Gail J. Brown, Randy Knize, and Yalin Lu. "Strain-induced high-temperature perovskite ferromagnetic insulator" PNAS **115**, 2873 (2018)

[26] Thornton, G, B C Tofield, and A W Hewat. "Study of LaCoO$_3$ in the Temperature Range 4.2< T < 1248 K." Journal of Solid State Chemistry **307**, 301 (1986)

[27] Frazer, Bradley H., Benjamin Gilbert, Brandon R. Sonderegger, and Gelsomina De Stasio. "The Probing Depth of Total Electron Yield in the Sub-KeV Range: TEY-XAS and X-PEEM." Surface Science **537**, 161 (2003)

[28] Liu, Boyang, Cinthia Piamonteze, Mario Ulises Delgado-Jaime, Ru-Pan Wang, Jakoba Heidler, Jan Dreiser, Rajesh Chopdekar, Frithjof Nolting, and Frank M. F. de Groot, "Sum rule distortions in fluorescence-yield x-ray magnetic circular dichroism." Physical Review B**96**, 054446. (2017)

[29] F. de Groot, "Multiplet effects in X-ray spectroscopy", Coordination Chemistry Reviews **249**, 31 (2005)

[30] Eva Grieten, Olivier Schalm, Pieter Tack, Stephen Bauters, Patrick Storme, Nicolas Gauquelin, Joost Caen, Alessandro Patelli, Laszlo Vincze, and Dominique Schryvers, "Reclaiming the image of daguerreotypes: Characterization of the corroded surface before and after atmospheric plasma treatment," Journal of Cultural Heritage **28**, 56–64 (2017).

[31] Bert Conings, Simon A Bretschneider, Aslihan Babayigit, Nicolas Gauquelin, Ilaria Cardinaletti, Jean Manca, Jo Verbeeck, Henry J Snaith, and Hans-Gerd Boyen, "Structure–Property Relations of Methylamine Vapor Treated Hybrid Perovskite CH$_3$NH$_3$PbI$_3$ Films and Solar Cells," ACS Applied Materials & Interfaces **9**, 8092–8099 (2017)

[32] DJ Groenendijk, C Autieri, T. C. van Thiel, W. Brzezicki, N. Gauquelin, P. Barone, K. H. W. van den Bos, S. van Aert, J. Verbeeck, A. Filippetti, S. Picozzi, M. Cuoco, A. D. Caviglia, "Berry phase engineering at oxide interfaces" - arXiv preprint arXiv:1810.05619, 2018.

[33] Vasiliev, A. N., Volkova, O. S., Lobanovskii, L. S., Troyanchuk, I. O., Hu, Z., Tjeng, L. H., Khomskii, D. I., Lin, H. J. Chen, C. T., Tristan, N., Kretzschmar, F., Klingeler, R. and Büchner, B., "Valence states and metamagnetic phase transition in partially B -site-disordered perovskite EuMn$_{0.5}$Co$_{0.5}$O$_3$." Physical Review B**77**, 104442 (2008)

[34] We point out that the very tiny shoulder at the lowest energy in the blue trace of Fig. 2(a) could signal a non-zero divalent contribution, but this is too small to fit reliably. We estimate the divalent contribution to be less than 3%.

[35] In the following, we refer to the process of forming divalent Co in the LaCoO$_3$ as 'charge transfer'. That this is the correct interpretation will become clear, and thus the readers' indulgence with this terminology at this point is appreciated.

[36] The 4-2-4 data are the same as those shown in Fig. 2[a].

[37] Ohtomo, A., D. A. Muller, J. L. Grazul, and H. Y. Hwang. "Epitaxial Growth and Electronic Structure of LaTiO$_x$ Films." Applied Physics Letters **80**, 3922 (2002)

[38] Liu, Guiju, Xiaotian Li, Yiqian Wang, Wenshuang Liang, Bin Liu, Honglei Feng, Huaiwen Yang, Jing Zhang, and Jirong Sun. "Nanoscale Domains of Ordered Oxygen-Vacancies in LaCoO$_3$ Films." Applied Surface Science **425** 121 (2017)

[39] V.V. Mehta, N. Biškup, C. Jenkins, E. Arenholz, M. Varela, Y. Suzuki, "Long-range ferromagnetic order in LaCoO$_3$ epitaxial films due to the interplay of epitaxial strain and oxygen vacancy ordering", Physical Review B **91**, 144418 (2015)

[40] Guiju Liu, Xiaotian Li, Yiqian Wang, Wenshuang Liang, Bin Liu, Honglei Feng, Huaiwen Yang, Jing Zhang, Jirong Sun, "Nanoscale domains or ordered oxygen-vacancies in LaCoO$_3$ films", Applied Surface Science **425**, 121 (2017)

[41] Ji-Hwan Kwon, Woo Seok Choi, Young-Kyun Kwon, Ranju Jung, Jian-Min Zuo, Ho Nyung Lee, Miyoung Kim, "Nanoscale Spin-state Ordering in LaCoO$_3$ Epitaxial Thin films", Chem. Mater. **26**, 2496 (2014)

[42] Woo Seok Choi, Ji-Hwan Kwon, Hyoungjeen Jeen, Jorge E. Hamann-Borrero, Abdullah Radi, Sebastian Macke, Ronny Sutarto, Feizhou He, George A. Sawatzky, Vladimir Hinkov, Miyoung Kim, Ho Nyung Lee, "Strain-Induced Spin States in Atomically Ordered Cobaltites", Nano Lett. **12**, 4966 (2012)

[43] Neven Biškup, Juan Salafrance, Virat Mehta, Mark. P. Oxley, Yuri Suzuki, Stephen J. Pennycook, Sokrates T. Pantelides, Maria Varela, "Insulating Ferromagnetic LaCoO$_{3-\delta}$ Films: A Phase Induced by Ordering of Oxygen Vacancies", Physical Review Letters **112** 087202 (2014)

[44] Chen, Yan, Dillon D. Fong, F. William Herbert, Julien Rault, Jean-Pascal Rueff, Nikolai Tsvetkov, and Bilge Yildiz. "Modified Oxygen Defect Chemistry at Transition Metal Oxide Heterostructures Probed by Hard X-Ray Photoelectron Spectroscopy and X-Ray Diffraction." Chemistry of Materials 30 3359 (2018)

[45] T. Saitoh, T. Mizokawa, A. Fujimori, M. Abbate, Y. Takeda, and M. Takano, "Electronic structure and temperature-induced paramagnetism in LaCoO$_3$", Phys. Rev. **B55**, 4257 (1997)

[46] Biesinger, M. C., Payne, B. P., Grosvenor, A. P., Lau, L. W. M., Gerson, A. R., & Smart, R. S. C., "Resolving surface chemical states in XPS analysis of first row transition metals, oxides and hydroxides: Cr, Mn, Fe, Co and Ni", Applied Surface Science, **257** 2717 (2011)

[47] G. van der Laan, C. Westra and G.A. Sawatzky, "Satellite structure in photoelectron and Auger spectra of copper dihalides." Phys Rev **B23**, 4369 (1981)

[48] Vaz, C. A.F., D. Prabhakaran, E. I. Altman, and V. E. Henrich, "Experimental Study of the Interfacial Cobalt Oxide in Co$_3$O$_4$/α-Al$_2$O$_3$ (0001) Epitaxial Films", Physical Review **B80**, 155457 (2009)

[49] Kim, K. S. "X-Ray-Photoelectron Spectroscopic Studies of the Electronic Structure of CoO." Physical Review **B11**, 2177 (1975)

[50] Thole, B. T., G. Van Der Laan, and G. A. Sawatzky. "Strong Magnetic Dichroism Predicted in the M$_{4,5}$ X-Ray Absorption Spectra of Magnetic Rare-Earth Materials." Physical Review Letters **55** 2086 (1985)

[51] Schütz, G., W. Wagner, W. Wilhelm, P. Kienle, R. Zeller, R. Frahm, and G. Materlik. "Absorption of Circularly Polarized x Rays in Iron." Physical Review Letters **58**, 737 (1987)

[52] Thole, B, P Carra, F Sette, and G Van der Laan. "X-Ray Circular Dichroism as a Probe of Orbital Magnetization." Physical Review Letters **68**, 1943 (1992)

[53] Csiszar, S. I., M. W. Haverkort, Z. Hu, A. Tanaka, H. H. Hsieh, H.-J. Lin, C. T. Chen, T. Hibma, and L. H. Tjeng. "Controlling Orbital Moment and Spin Orientation in CoO Layers by Strain." Physical Review Letters **95**, 187205 (2005)

[54] Jo, T., & Shishidou, T. "Orbital Magnetic Moments of CoO and FeO and Isotropic Co and Fe L$_{2,3}$ Absorption Spectroscopy." Journal of the Physical Society of Japan, **67**, 2505–2509 (1998)

[55] S. Blundell, Magnetism in Condensed Matter, Oxford University Press, (2001)

[56] The photon energy dependence of the L$_3$ XMCD signal for LCO4 shown in Fig. 6(a) is different to that of the samples with interfaces (Fig. 6[b-d]), an additional argument on top of the valence fingerprint data that the spins in LCO4 are not originating from Co[II]-3d$^7$ centres.

[57] Chang, C. F., Z. Hu, Hua Wu, T. Burnus, N. Hollmann, M. Benomar, T. Lorenz, et al. "Spin Blockade, Orbital Occupation, and Charge Ordering in La$_{1.5}$Sr$_{0.5}$CoO$_4$" Physical Review Letters **102**, 116401 (2009)

[58] Shannon, R D. "Revised Effective Ionic Radii and Systematic Studies of Interatomic Distances in Halides and Chalcogenides." Acta Crystallographica Section A **32**, 751 (1976)

[59] Thick LCO films grown on STO ($d_{LCO}$=60nm) using similar conditions to those given in Table I did exhibit ferromagnetism, observable both using VSM and element-specific (XMCD) magnetometry, in keeping with the literature for thicker $LaCoO_3$ films. The present (ultra)thin cobalt oxide layers may simply be below the minimal thickness required to enable FM ordering, as has been reported for PLD-grown $LaMnO_3$ films in Renshaw X. Wang et al., "Imaging and control of ferromagnetism in LaMnO3/SrTiO3 heterostructures." Science **349**, 716 (2015).

[60] Barla, A., G. Schmerber, E. Beaurepaire, A. Dinia, H. Bieber, S. Colis, F. Scheurer, et al. "Paramagnetism of the Co Sublattice in Ferromagnetic $Zn_{1-x}Co_xO$ Films." Physical Review B **76** 125201 (2007)

[61] Merz, M., P. Nagel, C. Pinta, A. Samartsev, H. v. Löhneysen, M. Wissinger, S. Uebe, A. Assmann, D. Fuchs, and S. Schuppler. "X-Ray Absorption and Magnetic Circular Dichroism of $LaCoO_3$, $La_{0.7}Ce_{0.3}CoO_3$ and $La_{0.7}Sr_{0.3}CoO_3$ Films: Evidence for Cobalt-Valence-Dependent Magnetism." Physical Review B **82**, 174416 (2010)

[62] Laan, Gerrit van der, Elke Arenholz, Rajesh V. Chopdekar, and Yuri Suzuki. "Influence of Crystal Field on Anisotropic X-Ray Magnetic Linear Dichroism at the $Co^{2+}$ $L_{2,3}$ edges." Physical Review B **77**, 064407 (2008)

[63] Maignan, A., V. Caignaert, B. Raveau, D. Khomskii, and G. Sawatzky. "Thermoelectric Power of $HoBaCo_2O_{5.5}$: Possible Evidence of the Spin Blockade in Cobaltites." Physical Review Letters **93**, 026401 (2004)

# Full control of Co valence in isopolar LaCoO$_3$ / LaTiO$_3$ perovskite heterostructures via interfacial engineering

Georgios Araizi-Kanoutas[1,*,a], Jaap Geessinck[2,*], Nicolas Gauquelin[3], Steef Smit[1], Xanthe Verbeek[1], Shrawan K. Mishra[4], Peter Bencok[5], Christoph Schlueter[6], Tien-Lin Lee[5], Dileep Krishnan[3], Jo Verbeeck[3], Guus Rijnders[2], Gertjan Koster[2] and Mark S. Golden[1,b]

[1]Van der Waals-Zeeman Institute for Experimental Physics, Institute of Physics, University of Amsterdam, Science Park 904, 1098 XH Amsterdam, The Netherlands

[2]MESA+ Institute for Nanotechnology, University of Twente, Faculty of Science and Technology, P.O. Box 217, 7500 AE Enschede, The Netherlands

[3]Electron Microscopy for Materials Science, University of Antwerp, Campus Groenenborger Groenenborgerlaan 171, 2020 Antwerpen, Belgium

[4]School of Materials Science & Technology, Indian Institute of Technology (BHU), Varanasi-221 005, India

[5]Diamond Light Source Ltd, Diamond House, Harwell Science & Innovation Campus, Didcot, OX11 0DE, United Kingdom

[6]PETRA III, DESY Photon Science, Notkestr. 85, 22607 HAMBURG, Germany

*These two authors contributed equally to this work.
[a] G.AraiziKanoutas@uva.nl   [b] M.S.Golden@uva.nl

## Supplementary material

### 1. Data treatment of XAS spectra

The XAS spectra were collected using simultaneously TEY and FY modes. In a few cases, extrinsic, but well understood peaks made their appearance in the spectra. Firstly, the TEY signal of Co L-edge exhibited weak Ba-M$_{4,5}$ absorption the origin of which is elusive. This signal did not give rise to any dichroism and subtraction was achieved by measuring the Co L-edge of samples that did not contain any Co (namely LTO 4 u.c.), as shown in Figure S1a. Secondly, the FY signal, used primarily to ensure that TEY is sufficiently representative of the bulk of the film, contained an Al:Kα peak that originated from 2$^{nd}$ order synchrotron radiation exciting the K-edge of Al contained in the LAO buffer layer. The same Co L-edge of a reference film containing no Co was used to subtract this feature, too. Further investigation showed that the SiC coated mirror is adequate to strongly suppress this peak in contrast to the Au mirror.

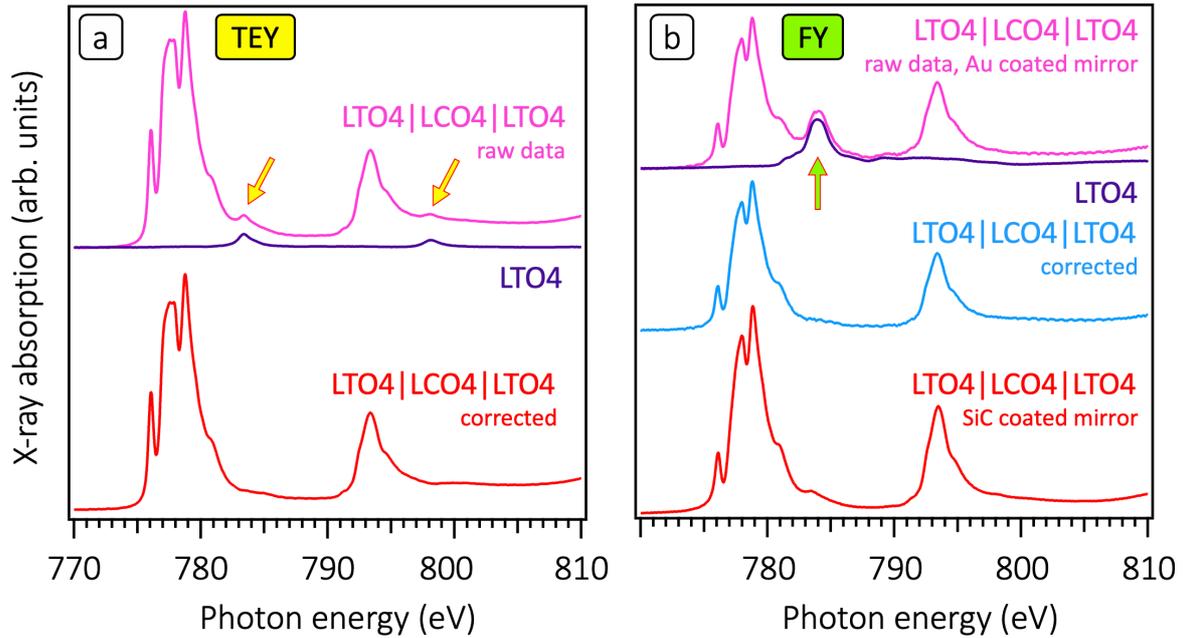

**Figure S1: Correction of Ba-$M_{4,5}$ signals in TEY and Al:K$\alpha$ fluorescence in FY.** (a) TEY mode XAS data show very weak Ba-$M_{4,5}$ signals in the energy region of the Co-$L_{2,3}$ edge which are unrelated to the LCO layers. In the main figures, the Co-free, 4 u.c. LTO data are used to subtract this signal. (b) The FY data do not contain the Ba-$M_{4,5}$ features, but do show Al:K$\alpha$ fluorescence excited by second order light. Again, the LTO spectrum is used to correct the data. In the lowest trace, a beamline mirror was used which does not transmit the $2^{nd}$ order light, thus offering a clear control of the correction.

## 2. Co valence fingerprinting using model compounds

Two methods were explored to determine the % of divalent Co 3d7 in the [0*IF], [1*IF], and [2*IF] structures. In a first iteration, the spectra were modelled using the data from single crystalline model compounds from Ref. [1]. An example of fits is shown in Fig. S2(a), with the percentages as given in the right-hand panel. The heterostructure spectra are quite well described in this manner, and the 'stand-alone' LCO film shows more intensity at 778eV suggesting (in the light of Fig. 2(b) in the main body of the paper) that the thin-film LCO has a greater HS admixture than the single crystal. We note the divalent pre-peak at 777eV is clearly a different feature to this shoulder, as can clearly be seen in the heterointerface samples.

In atomic multiplet-based models for $L_{2,3}$ XAS of transition metal oxides, the exact spectral distribution of multiplets in Co-$L_{2,3}$ XAS (which ends up giving the shape of the overall spectrum) is sensitive to a number of fundamental parameters including Coulomb energies (dd and core-d), the charge transfer energy, covalence between the metal and oxygen levels as well as crystal fields and the local crystal symmetry.

It is asking a lot to expect the finer details of the multiplets from the bulk, model compounds to be strictly relevant for the ultrathin film systems here. For the divalent cobalt case, CoO is quite a different system structurally, to a LCO4|LCO4|LTO4 trilayer, grown coherently strained to STO. Therefore, in a second iteration, the series of Co-$L_{2,3}$ spectra were decomposed into tri- and divalent contribution using the Co $L_{-2,3}$ XAS traces from LCO4 and the LCO4|LCO4|LTO4 trilayer as representatives of trivalent and divalent Co in the structural context of our thin film heterostructures. Fig. S2(b) shows the results of this process, which yield excellent agreement with the details of the multiplet structures. The numbers shown in Fig. S2(b) are given in Table II in the main body of the paper.

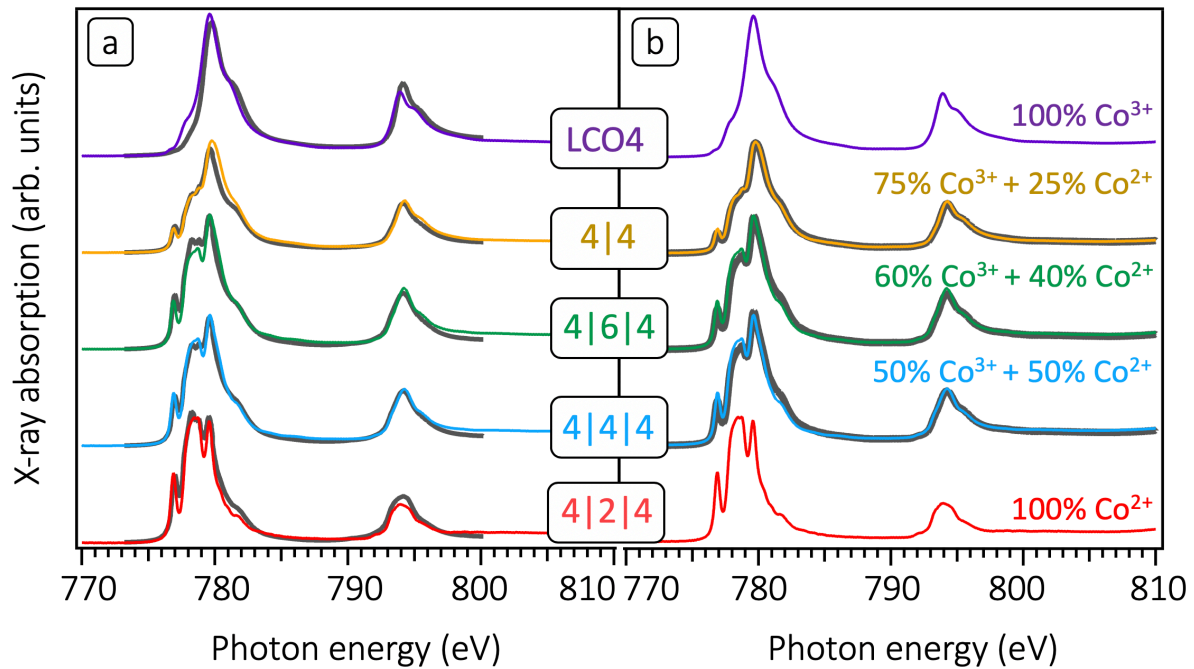

Figure S2: Decomposition of the Co-$L_{2,3}$ XAS data to extract tri- and divalent contributions. (a) Use of the spectra from the model compounds $EuCoO_3$ and CoO to fit the data from bi- and trilayer heterointerface samples. The same mixing percentages of the di- and trivalent model spectra are kept as used in the right-hand panel. (b) An analogous procedure – the one actually used to determine the valency figures given in the main body of the paper – in which the data from the LCO4 and LTO4|LCO2|LTO4 systems are used a tri- and divalent model spectra, respectively, mixed as indicated in the figure.

### 3. Ti 3d electron count from Ti-$L_{2,3}$ XAS of LTO|LCO|LTO trilayers

The valency of the Ti ions in the LTO layers in the investigated heterostructures was probed using both Ti2p HAXPES at the I09 beamline, and soft X-ray XAS at the Ti-$L_{2,3}$ edges at both the I09 and the I10 beamlines. The data from the I10 experiments are shown in Fig. S3 from [2*IF] LTO4|LCO2|LTO4 (424), 444 and 464 samples. Due to charge transfer to LCO, the single 3d electron present in stoichiometric, defect-free LTO should be missing, yielding a $Ti^{4+}$ XAS spectrum, as seen in Fig. S3. As discussed in the sample design section of the main paper and in the context of Fig. S3 below, even though the growth conditions were carefully chosen to optimally grow both LCO and LTO with as few defects and vacancies as possible, in the post-growth, post-cool-down timepoint, a single film of LTO does not necessarily possess pure $3d^1$ character for reasons explained in Ref. [2]. We reiterate the message of the main paper here that the essential comparison is between the valence states of the Ti in LTO abutting LCO and the Co in LCO abutting LTO between the situation in experiment, and – for example - the DFT calculations of Ref. [3]. It is evident that the situation of a Ti/Co interface possessing $3d^1/3d^6$ (i.e. tri/divalent valencies) is avoided, both in the computer and in practise.

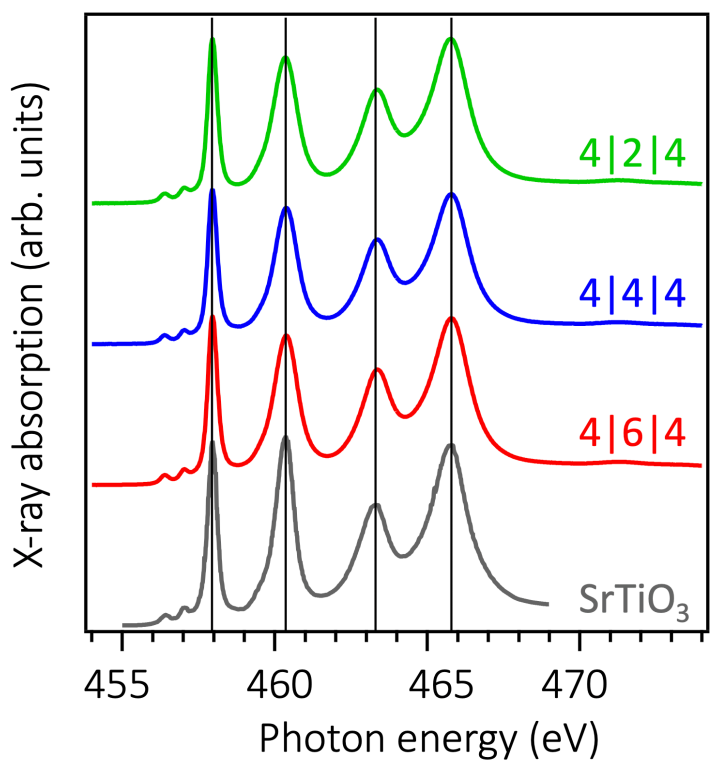

Figure S3: Ti-$L_{2,3}$ XAS data from [2*IF] samples and STO reference. Shown are TEY data for 2 (green), 4 (blue) and 6 (red) uc of LCO sandwiched between a pair of 4 uc thick LTO layers. Their strong resemblance to the $Ti^{4+}$ $3d^0$ data from a STO reference from the literature [2] indicates a tetravalent configuration.

4. Control of the Ti 3d electron count in LTO films through UHV heat treatment

Figure S4 shows in-situ Ti2p core level photoemission data recorded from a LAO 3u.c./LTO 3 u.c./LAO 3u.c. film grown on LAO substrate in using identical growth parameters to those of the LTO layers in the bi- and tri-layer systems that form the core of this paper. The Ti2p spectral shape of the as-grown film (blue) yields a $d^0$ character due to over-oxidation that took place in the relatively high growth pressure environment that was required to enable growth of LCO/LTO film combinations in agreement to recent results [2]. Subsequently the film was transferred in-situ to the XPS chamber in a pressure environment of $1\times10^{-10}$mbar. Having undergone thermal post-treatment in ultrahigh vacuum and at temperatures comparable to the deposition conditions, the LTO shows increased $3d^1$ content expressed in the peak located at the low binding of each spin-orbit split component of the doublet. Still higher annealing temperatures further promote the stabilisation of the $3d^1$ configuration. After cooldown (uppermost trace in Fig. S4[a]), the film exhibits a dominant $Ti^{3+}$ population, reflecting establishment of the relatively unstable stoichiometric state of LTO. In the right-hand panel of the figure, the data show how consecutive cycles of heating and cooling yield the same spectral signatures, highlighting the tendency of LTO to gain oxygen once at elevated temperature and in relatively higher pressures ($10^{-9}$mbar at 700$^o$C, compared to $10^{-10}$mbar at room temperature). Restoration of the dominance of the stoichiometric valence was again achieved upon subsequent cooling.

From these data it is clear that there is a reversible ingress/removal of additional oxygen from the LTO within the depth-scale of the XPS measurements.

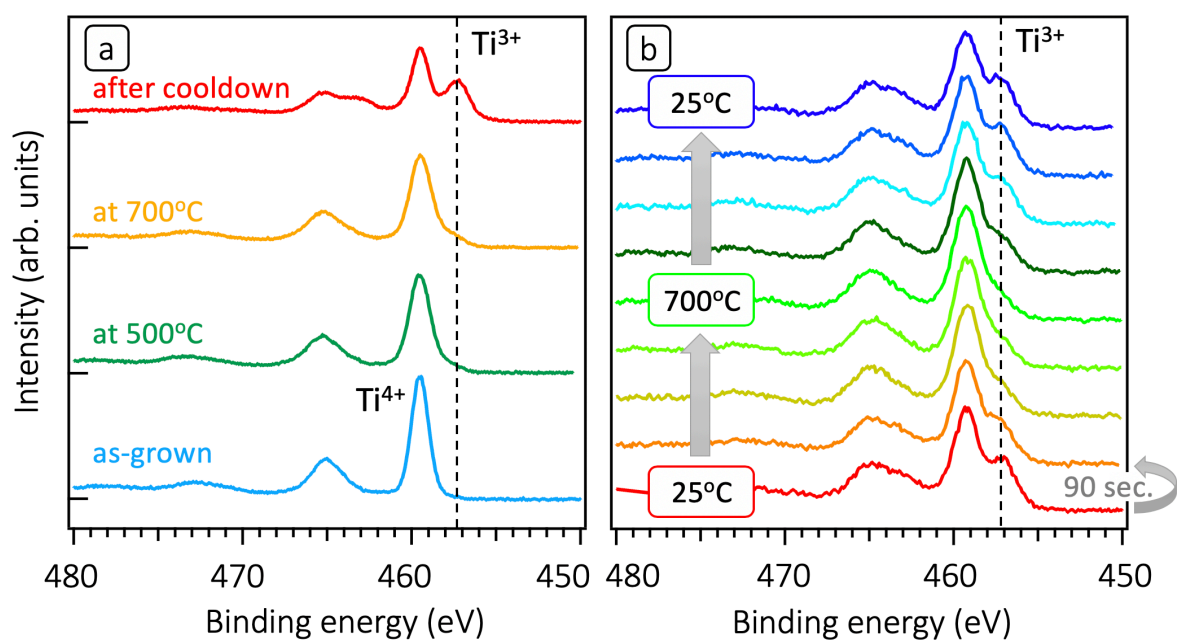

Figure S4: Ti-2p XPS data from a LTO film grown on LAO. The LTO is 4 uc thick, deposited on a 3 uc cell LAO layer itself grown on bulk LAO The film stack of capped with a further 3 uc of LAO to yield $LAO_{bulk}|LAO3|LTO4|LAO3$. Panel (a) on the left shows data after cool-down, transferred without breaching UHV (lowest trace), and at different post-anneal temperatures as indicated and after a final cool-down from 700°C. The characteristic $2p^5 3d^1$ final state feature for $Ti^{3+}$ is clearly present at low binding energies in the latter. Panel (b) on the right shows reversible behaviour during a heating/cooling cycle between 25 and 700°C.

We mention that the subjection of a LTO4|LCO4|LTO4 trilayer, grown on LAO and capped with LAO - i.e. analogous to the LTO sample whose data is shown in Fig. S4 - to in-situ annealing steps in UHV starting at 25°C and going via 175°C up to 400°C (i.e. all below the annealing temperatures used in Fig. S4 for LTO) led to the appearance of a $Ti^{3+}$ feature in XPS signaling 3d electron occupation in the LTO but also an XPS signal at a binding energy matching cobalt metal, signaling the irreversible decomposition of the LCO under these conditions in the trilayer. These experiments support the chosen strategy of growth at an intermediate oxygen pressure.

## 5. Raw XAS spectra using circular polarised X-rays and the resulting XMCD contrast

Figure S5 shows an example of the raw, polarisation-dependent XAS data from which the XMCD datasets shown in Fig. 7 of the main paper were extracted.

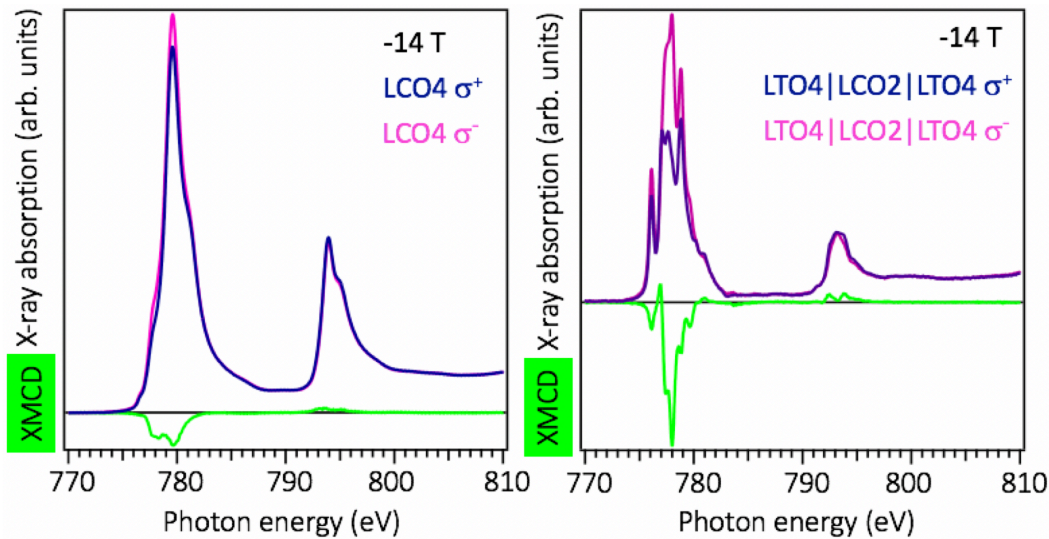

Figure S5: Example X-ray absorption spectra used to extract XMCD data. Co-$L_{2,3}$ XAS data (TEY) for an LTO4 (left) and LTO4|LCO2|LTO4 sample (right), together with the XMCD signal (in green) as shown for different film systems in Fig. 7 of the main paper.

## 6. Temperature- and incidence angle dependent XMCD of LTO|LCO|LTO trilayers

Figure S6(a) shows the XMCD data of the double-interface system LTO4|LCO6|LTO4 measured at a field of -14 T at different temperatures. These data form the basis of the data points shown in the inset of Fig. 7 in the main paper. In Fig. S6(b), a comparison between the grazing incidence (NI) and normal incidence (NI) XMCD data for the double-interface system LTO4|LCO4|LTO4 is presented, showing only very small deviations in the XMCD patterns despite probing magnetic polarisation either in (GI) or out of the film plane (NI).

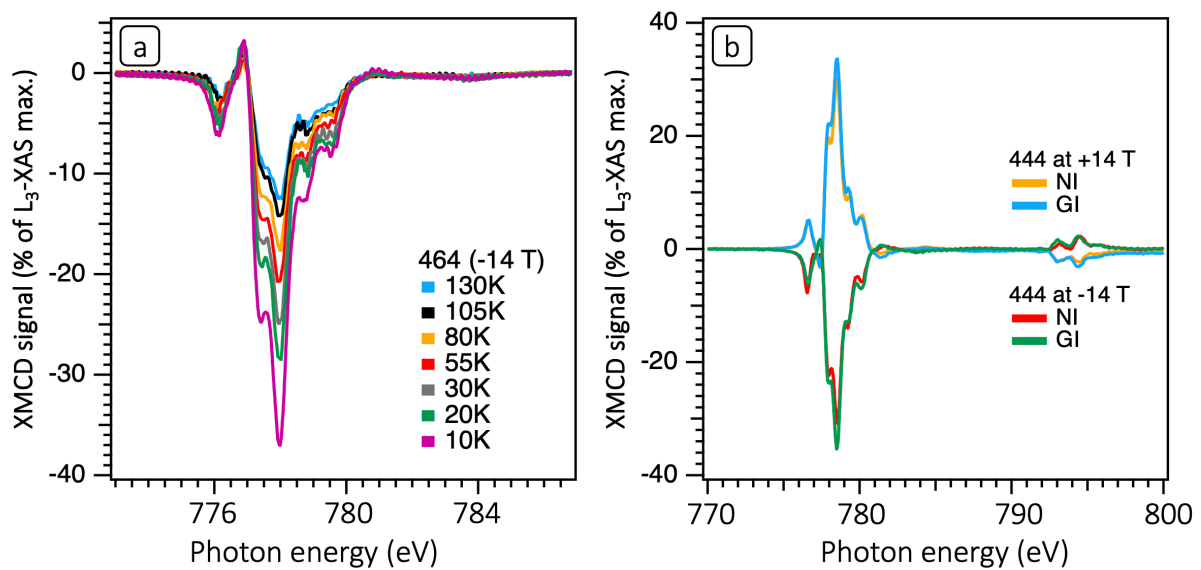

Figure S6: Temperature and angular signatures of paramagnetic behaviour.
Panel (a) Temperature dependence of the system LTO4|LCO6|LTO4 XMCD signal from measured at a field of -14T. These data underpin the data-points of Fig. 8 in the main paper.

(b) Normal incidence (NI) vs. grazing incidence (GI) XMCD patterns of system LTO4|LCO4|LTO4 measured at both at 14 and -14 Tesla fields. Essentially zero anisotropy is observed.

# References

[1] Vasiliev, A. N., Volkova, O. S., Lobanovskii, L. S., Troyanchuk, I. O., Hu, Z., Tjeng, L. H., Khomskii, D. I., Lin, H. J. Chen, C. T., Tristan, N., Kretzschmar, F., Klingeler, R. and Büchner, B., "Valence states and metamagnetic phase transition in partially B -site-disordered perovskite $EuMn_{0.5}Co_{0.5}O_3$." Physical Review **B77**, 104442 (2008)

[2] Philipp Scheiderer, Matthias Schmitt, Judith Gabel, Michael Zapf, Martin Stübinger, Philipp Schütz, Lenart Dudy, Christoph Schlueter, Tien-Lin Lee, Michael Sing, and Ralph Claessen "Tailoring Materials for Mottronics: Excess Oxygen Doping of a Prototypical Mott Insulator" Adv. Matter. **30** 1706708 (2018)

[3] Zhong, Zhicheng, and Philipp Hansmann. "Band Alignment and Charge Transfer in Complex Oxide Interfaces." Physical Review X **7** 011023 (2017)